\documentclass [aps,prL,amssymb,amsmath,twocolumn,showpacs]{revtex4-1}

\usepackage{microtype}
\usepackage{amsmath}
\usepackage{pifont}
\usepackage{amssymb}
\usepackage{amsthm}
\usepackage{mathrsfs}
\usepackage{graphicx}
\usepackage{verbatim}
\usepackage{subfigure}
\usepackage{float}
\usepackage{bm}
\usepackage{booktabs}
\usepackage{multirow}

\usepackage[dvipsnames]{xcolor}
\usepackage{graphicx,epsfig,amsfonts,amssymb}
\usepackage{hyperref}
  \hypersetup{colorlinks=true,citecolor=blue}

\usepackage{bm}% bold math
\usepackage{times}
\usepackage{lipsum}

\usepackage[all]{xy}

\newcommand{\nn}{\nonumber}

\newcommand{\br}{\mathbf r}

\newcommand{\be}{\begin{eqnarray}}
\newcommand{\ee}{\end{eqnarray}}
\newcommand{\la}{\langle}
\newcommand{\ra}{\rangle}
\newcommand{\rar}{\rightarrow}

\begin{document}
	
\title{Kibble-Zurek Behavior in the Boundary-obstructed Phase Transitions}
\author {Menghua Deng$^1$}
\author {Zhoujian Sun$^1$}
%\author { Adolfo del Campo$^{2,3}$}

\author{Fuxiang Li$^1$}
%\email{fuxiangli@hnu.edu.cn}
\affiliation{$^1$School of Physics and Electronics, Hunan University, Changsha 410082, China}
%\affiliation{$^2$ Department of Physics and Materials Science, University of Luxembourg, L-1511 Luxembourg, Luxembourg}
%\affiliation{$^3$ Donostia International Physics Center, E-20018 San Sebastián, Spain}

\date{\today}

\begin{abstract}
 We study the nonadiabatic dynamics of  a  two-dimensional higher-order  topological insulator when the system is slowly quenched across the boundary-obstructed phase transition, which is characterized by edge band gap closing.   
 We find that the number of excitations produced after the quench exhibits power-law scaling behaviors with the quench rate. Boundary conditions can drastically modify the scaling behaviors: The scaling exponent is found to be $\alpha=1/2$ for hybridized and fully open boundary conditions, and  $\alpha=2$  for periodic boundary condition.  We argue that the exponent $\alpha=1/2$ cannot be explained by the Kibble-Zurek mechanism unless we adopt an effective dimension $d^{\rm eff}=1$ instead of the real dimension $d=2$.  For comparison, we also investigate the slow quench dynamics across the bulk-obstructed phase transitions and a single multicritical point,  which obeys the Kibble-Zurek mechanism with dimension $d=2$. %The results of quenches across the latter two types of phase transitions all obey the {\color{red} Kibble-Zurek scaling with dimension $d=2$, but have different scaling exponent}.
 \end{abstract}

%\pacs{05.60.-k, 05.40.-a, 82.37.-j, 82.20.-w}
%05.60.-k Transport processes
%05.40.-a Fluctuation phenomena, random processes, noise, and Brownian motion
%05.70.Ln nonequilibrium thermodynamics
%05.10.Gg Stochastic analysis methods (Fokker-Planck, Langevin, etc.)
%62.25.Fg High-frequency properties, responses to resonant or transient (time-dependent) fields
%82.20.-w Chemical kinetics and dynamics
%82.37.-j Single molecule kinetics

\date{\today}

\maketitle

%%%%%%%%%%%%%%%%%%%%%%%%%%%%%%%%%%%

%\begin{figure} [!htb]
%{\includegraphics[width=2\columnwidth]{fig1.png}}
%\caption{Adiabatic spectrum for BCS Hamiltonian (\ref{h-bcs0}) for different couplings $g=1/N_s$ (a) and $g=1$ (b). Total number of spins is $N_s=10$. The energy levels are $\ve_j=1-j/N_s$. } \label{fig:spectrum}
%\end{figure}
%

{\it Introduction.}--
As one of the cornerstones in nonequilibrium physics,  the Kibble-Zurek (KZ) mechanism  plays an important role in the understanding of both classical continuous phase transition and quantum phase transition \cite{Kibble1976,Kibble1980,Zurek1985,Zurek1996, Campo2014}, and in the applications from condensed matter physics to adiabatic quantum computing \cite{Damski2005,Zurek2005,Polkovnikov2005,Sen2008,Barankov2008,Dziarmaga2005, Dziarmaga2010,Polkovnikov2011,Nowak2021,Kou2023,Liang2024}. 
 According to KZ mechanism, when a system is slowly quenched across a critical phase transition point,  topological excitations are formed and the density of these excitations $n_{exc}$ demonstrates a universal power-law scaling relationship with the quenching  rate $R$:  
\be
n_{exc} \sim R^{\alpha}~~ {\rm with} ~~ \alpha=\frac{d\nu}{1+z\nu}. \label{eq1}
\ee
 Here, $d$ is the system dimension, and $\nu$ and $z$ are the exponent of correlation length and dynamical critical exponent, respectively. KZ mechanism was successfully tested by numerical simulation \cite{Languna1997,Yeates1998,Dziarmaga1999,Antunes1999,Bettencounrt2000,Uhlmann2007,Witkowska2011,Das2012,Sonner2015,Chesler2015} and experimentally verified in a variety of platforms \cite{Deutschlander, Maegochi, Du2023, Weiler, Lamporesi, Navon, Anquez, Ko2019, Yi2020, Keesling, Ebadi}. 
It is remarkable that this mechanism is universal and not affected by local perturbations such as different boundary conditions \cite{Gomez2022}. Rapid progress has been made in this research area and the idea of KZM has been expanded  to  rapid quench regime, full counting statistics and large deviations \cite{Campo2018,Ruiz2019, Cui2020, Gomez2020, Bando2020,Mayo2021,Campo2021,Zeng2023, Balducci2023}. %Besides, the anti-Kibble-Zurek behavior has been studied in quantum systems that are coupled to environment or are in the presence of noise \cite{Dutta2016,Gao2017,Meier2017,Puebla2020,Singh2021,Ai2021,Singh2023}.

Topological insulators have attracted growing theoretical and experimental interest in the past two decade \cite{Kane2005,Kane200502,Bernevig2006,Qi2008,Moore2009,Franz2013,Hasan2010, Qi2011}. Topological phases are characterized by integer topological invariants defined over the bulk and by the existence of boundary states localized on the edge. The topological invariants cannot change their quantized value without closing the bulk energy gap at the topological phase transition point. 
Beyond the scope of Landau-Ginzburg-Wilson framework, topological phase transition may induce novel phenomena in nonadiabatic dynamics of a driven system.  Indeed, it has been suggested that the topology of the system may induce an anomalous defect production which strongly deviates from the KZ scaling law \cite{Bermudez2009,Bermudez2010,Liou2018}.  Despite these efforts, the intriguing physics induced by the interplay between  higher order topological phase transition and nonadiabatic dynamics has not been explored. 

In this Letter we investigate the Kibble-Zurek behavior after nonadiabatic slow quench across a  boundary-obstructed topological phase transition (BoOPT)  \cite{Khalaf2021} in a two-dimensional higher-order insulator \cite{Slager2015,Benalcazar2017,Benalcazar201702,Liu2017,Song2017, Langbehn2017,Ezawa2018,Khalaf2018,Geier2018,Schindler2018, Trifunovic2019,Benalcazar2019,Pozo2019,Zhang2019,Liu2019, Schindler2019,Li2020,Ezawa2020,Pererson2020,Xie2021,Tiwari2020, Asaga2020,Claes2020,Ghadimi2023,Benalcazar2022}. 
In contrast to the bulk-obstructed topological phases that demands bulk gap closings at the phase transition point, BoOTPs are characterized by the gap closing of edge states at the critical point. The boundary-obstructed topological phases have been predicted to  exist in ion-based topological superconductors \cite{Wu2020} and have been experimentally realized  in acoustic crystals \cite{Du2022}.

We find that the density of excitations  exhibits power-law scaling behaviors  with the quenching rate, as expected from KZ mechanism. However, the scaling exponents are drastically different for different boundary conditions: $\alpha = 2$ for  periodic boundary condition in both directions, while $\alpha=1/2$ for hybridized  (periodic in one direction, open in the other) and open boundary condition in both directions. %This result is the first example that boundary conditions can dramatically  modify  the KZ behavior. 
Moreover, the exponent $1/2$ cannot be explained by the KZ mechanism  if one takes the value of $d=2$ and $\nu=z=1$ in Eq.~(\ref{eq1}). Rather, one has to take into account the fact that in the BoOPT, the edge state plays the dominant role and thus the effective dimension should be taken as $d^{\rm eff}=1$.   This picture is further validated by analytical solution of an effective Landau-Zener model describing the dynamics of edge states and by the numerical calculation of real-space distribution of the excitations after quench. For comparison, we also study the nonadiabatic dynamics of the bulk-obstructed phase transition (BuOPT) and the isolated multicritical point (MCP). 
The scaling exponents are $\alpha=1$ and $0.455$, respectively, and agrees with KZ mechanism if we adopt the real dimension $d=2$ and exponents $z$ and $\nu$ obtained from finite-size scaling. These findings deepen our understanding of nonadiabatic quantum dynamics induced by topology and enriches the  Kibble-Zurek phenomena.

%Since the energy gap closing at BoOPTs is related to the boundary conditions (BCs), we consider slow quenches for different BCs in a two-dimensional higher-order topological insulators with BOTPs. The dynamics of the bulk-obstructed phase transitions (BuOPTs) and the isolated multicritical point (MCP) have been studied as comparisons. For each type of phase transition, we discuss three kinds of boundary conditions: periodic (periodic in both directions, PBCs), hybridized (periodic in one direction, open in the other, HBCs), and open (open in both directions, OBCs). In the case of BoOPT, the number of excitations for the system with PBCs scales with the quench rate $R$ as $R^4$. However, for both HBCs and OBCs, the number of excitations in the (relatively) slow-quench regime scales as $R^{1/2}$, which conforms to the KZ prediction. In the cases of the BuOPTs and MCP, the {\color{red}numbers} of the excitations scale with the quench rate $R$ as $R^1$ and $R^{0.455}$, respectively, which are independent of the BCs. The results of the BuOPTs and MCP agree with the KZ scaling but correspond to different critical exponents $z$ and $\nu$. We also calculate the real-space distribution of the excitations for the system with OBCs. The excitations produced after a quench across the BoOPT, are entirely localized at the edges of the system. For the quenches across the BuOPTs or MCP, the excitations, however, are present throughout space.  

\begin{figure}[t]
	\centering
	\epsfig{file=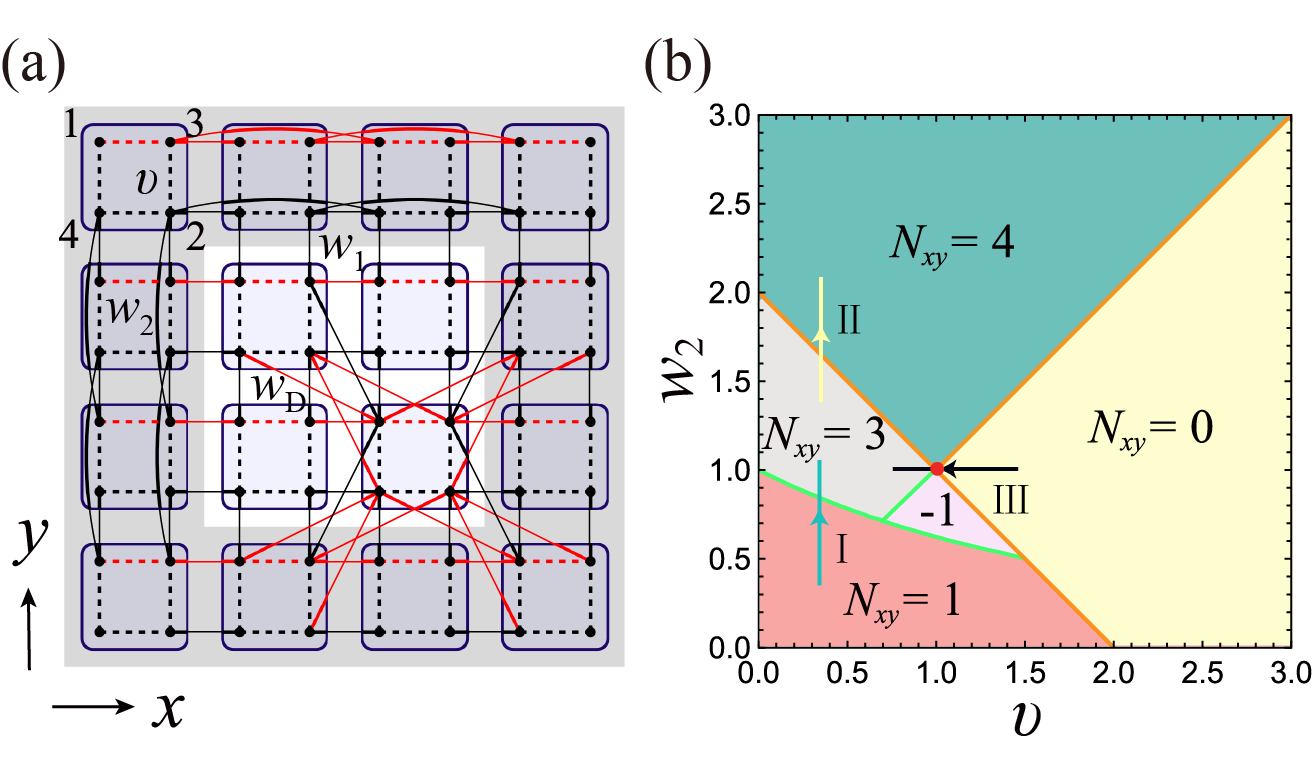, width=3.5in}
	\caption{(a) Schematics of the $C_{4\nu}$-symmetric model. Some non-nearest-neighbor hoppings are not shown for clarity. All red hoppings are multiplied by -1 to account for a flux of $\pi$ threading each plaquette. %Shaded area indicates the edges of the system. 
	(b) Phase diagram characterized by MCN $N_{xy}$. BoOPTs, BuOPTs and the isolated MCP are highlighted by  green line, orange line and red dot, respectively.   The arrows illustrate quenches across different kinds of phase transitions. Other parameters $w_1=1$, $w_D=0.5$ and $w_{m>2}=0$. }
	\label{fig:pd}
\end{figure}

{\it Model.}--Without loss of generality, we consider a chiral and long-range quadrupole topological insulator (QTI) as an example \cite{Benalcazar2017, Benalcazar2022}. The model consists of four sublattices in one unit cell, and includes   the nearest-neighbor intracell hopping $v$ and intercell hopping $w_1$, and the next-nearest-neighbor hopping $w_2$, see Fig.~\ref{fig:pd}(a). Besides, the hopping $w_D$ among nearest-neighbor unit cells along the diagonal directions is also needed, which preserves the chiral and $C_4$-symmetries but breaks the separability between the $k_x$ and $k_y$ momentum sectors. A $\pi$ flux is introduced to thread through each plaquette, and thus renders a negative sign of the hoppings as denoted by red dashed and solid line.
The corresponding Bloch Hamiltonian  reads 
\be\label{eq:hk1}
\mathcal{H}(\mathbf{k})=\mathbf{h}(\mathbf{k})\cdot\mathbf{\Gamma}=\sum_{i=1}^{4}h_i(\mathbf{k})\Gamma_i,
\ee
where $\Gamma_0=\tau_3\otimes\sigma_0,\ \Gamma_{k}=-\tau_1\otimes\sigma_{k},\ \Gamma_4=-\tau_2\otimes\sigma_0$ for $k=1,2,$ and $3$; $\tau, \sigma$ are Pauli matrices, and $\sigma_0$ is the $2 \times 2$ identity matrix. The $\mathbf{h}(\mathbf{k})$ is given by
	\be \label{eq:hk2}
	&&h_1(\mathbf{k})=-v_y-\sum_m^M w_{m,y}\cos(mk_y)+2w_D \cos k_x\cos k_y;\nn\\
	&&h_2(\mathbf{k})=\sum_m^M w_{m,y}\sin(mk_y)-2w_D \cos k_x\sin k_y.
\ee
$h_3$ and $h_4$ can be obtained from $-h_1$ and $h_2$, respectively,  by exchanging $k_x$ and $k_y$  and replacing $v_y$ by $v_x$.  Here, $M$ denotes the maximum number of long-range hoppings.
The explicit form of the Hamiltonian in the real space is provided in the Supplemental Material (SM)\cite{sm}.  

 %As we will show in the following, this model offers a rich scenario to study the influence of boundary conditions on the defect production across a critical point, and may serve as a paradigmatic model for more general topological systems.
It has been shown that in such a QTI model, there exists a novel kind of higher-order topological phases that are generally boundary-obstructed, and are identified by a $\mathbb{Z}$ topological invariant known as the multipole chiral numbers (MCNs) \cite{Benalcazar2022}. In a 2D system,  the MCNs is calculated as
\be\label{eq:z1}
N_{xy}=\frac{1}{2\pi i}{\rm Tr\, log}(\bar{\cal{Q}}^A_{xy}\bar{\mathcal{Q}}^{B\dagger}_{xy})\in\mathbb{Z},
\ee
where $\bar{\cal{Q}}^{\mathcal{S}}_{xy}=U^\dagger_{\mathcal{S}}\mathcal{Q}^{\mathcal{S}}_{xy}U_{\mathcal{S}}$ for sublattice $\mathcal{S}$=$A,B$.  $\mathcal{Q}_{xy}^{\mathcal{S}}=\sum_{\mathbf{r},\alpha\in\mathcal{S}}\vert\mathbf{r},\alpha\rangle \exp(-i{2\pi xy}/{L_{x}L_{y}})\langle\mathbf{r},\alpha\vert$ is the sublattice quadrupole moment operator with $\mathbf{r}=(x,y)$ labeling the unit cells. $U_{\mathcal{S}}$ is a unitary matrix representing the space spanned by $\{\psi^{\mathcal{S}}_n\}$, i.e., $U_{\mathcal{S}}=(\psi^{\mathcal{S}}_1,\psi^{\mathcal{S}}_2,...,\psi^{\mathcal{S}}_{N_{\mathcal{S}}})$, where $\psi_{\mathcal{S}}$ is normalized vector that exist only in the $\mathcal{S}$ subspace. 
 The phase diagram of the QTI model is shown in Fig.~\ref{fig:pd}(b). 
It is interesting to note that phase transitions between phases with different MCNs
need not close the bulk band gap but may close only the edge or surface band gap. As we will show in the following, this model offers a rich scenario to study the influence of boundary conditions on the defect production across a critical point, and may serve as a paradigmatic model for more general topological systems.%Indeed, one can see that there are three types of phase transitions in the phase diagram, BoOPTs (solid green line), BuOPTs (solid silver line), and an isolated MCP (the red point). Our work will show that these different types of phase transitions exhibit distinctive Kibble-Zurek behaviors under nonadiabatic quench dynamics, and under different boundary conditions. 

{\it Quenches.}--We initially prepare the system  in the ground state  with the valence band filled and the conduction band empty, and then slowly quench the system by  linearly varying  the parameters $v$ or $w_2$.
As denoted in Fig.~\ref{fig:pd}(b), we consider three different quench protocols corresponding to different types of phase transitions: (I) quench across the BoOPT with varying $w_2(t)$; (II)  quench across the BuOPT with varying  $w_2(t)$; and (III)  quench across the isolated MCP with varying $v(t)$. %We separately show that the power-law relation between the excitations and the quench rate, predicted by the KZ mechanism, arises after the quenches across the three types of phase transitions.

By choosing each valence band eigenstate of the pre-quench Hamiltonian as an initial state, one can solve the Schr$\mathsf{\ddot{o}}$dinger equation, and obtain the post-quench state $|\Phi_v(t_f)\rangle$.  The number of excitations is defined as the projection of $|\Phi_v(t_f)\rangle$ on the conduction band eigenstate $\vert\Psi_c(t_f)\rangle$ of post-quench Hamiltonian, $N_{exc}=\sum_{c, v}\vert\langle\Psi_c(t_f)\vert\Phi_\nu(t_f)\rangle\vert^2$. Here, the summation is taken over the valence and conduction bands. We will also need the real-space distribution of excitation defined as: $n_{exc}(\br)=\sum_{c, v}|\la \br|\Psi_c \ra|^2 \vert\langle\Psi_c\vert\Phi_\nu\rangle\vert^2$.

\begin{figure*}[htbp]
	\centering
	\epsfig{file=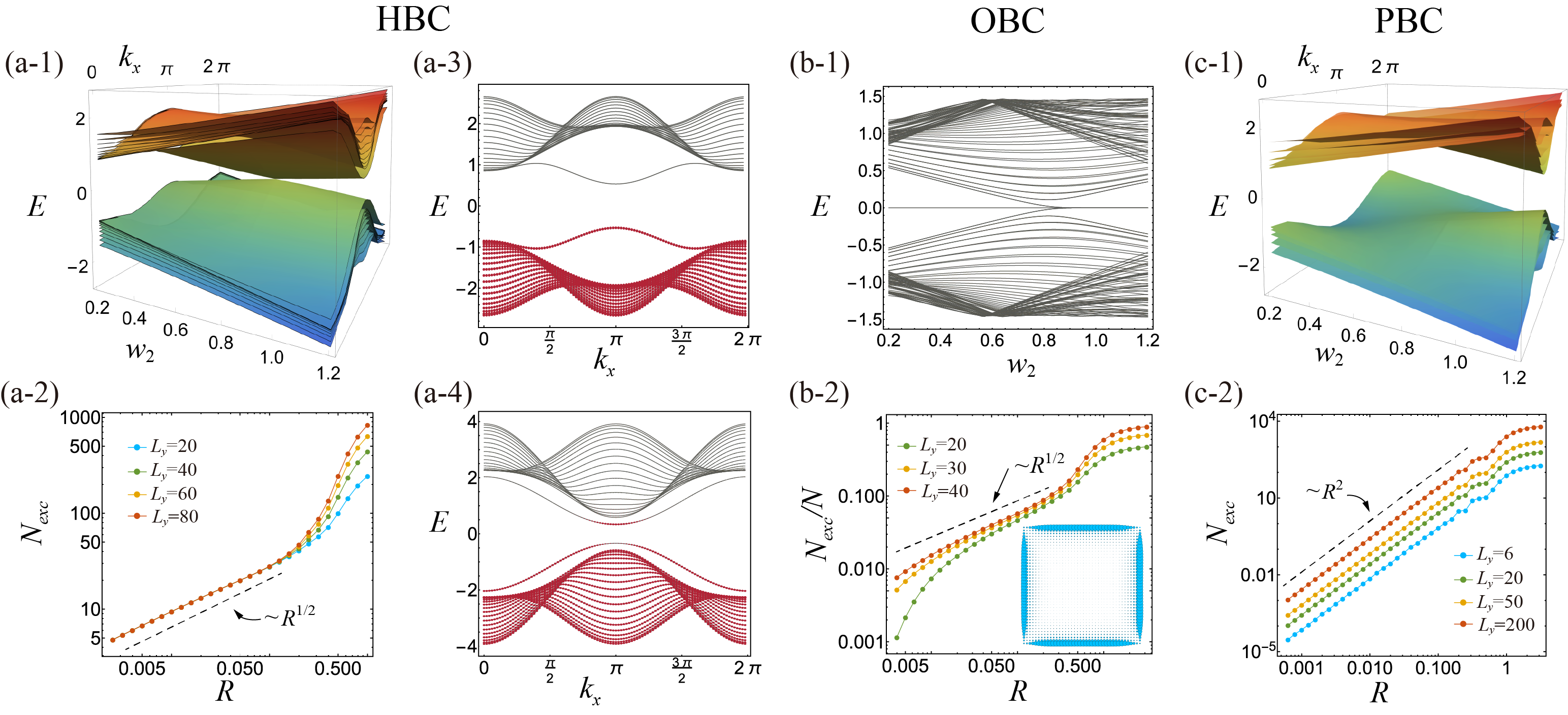, width=7in}
	\caption{Kibble-Zurek behaviors in the BoOPT for  different boundary conditions (a) HBC, (b) OBC and (c) PBC. Energy spectrum for different boundary conditions are plotted in (a-1), (b-1) and (c-1). 
Number of excitations $N_{exc}$ in the function of $R$ for different system widths $L_y$ are plotted in (a-2), (b-2) and (c-2), and the scaling exponents are 1/2, 1/2 and 2, respectively.  
	In (a-1), a pair of two-fold degenerate bands appear in the bulk-energy gap for HBC (comparison with (c-1)).  
	(a-3) and (a-4) plot the electron occupation of energy bands before and after the quench, respetively. The size of the dots is proportional to the occupation probability. Real-spatial distribution of excitations after quench with $R=10^{-1.2}$ is shown in the inset of  (b-2) . In (a) and (c), we consider strips with fixed large $L_x$ sites in the $x$ direction and  $L_y$ sites in the $y$ direction. In (b), we consider square lattices with $L_x=L_y=L$. %$N$ is the number of occupied states before the quench. 
	Parameters: $v=0.4$, $w_1=1$ and $w_D=0.5$; $w_2$ varies in $[0.2,1.2]$.  }
	\label{fig:Bo}
\end{figure*}

{\it Quench across BoOPT.}--We first study the quench dynamics across BoOPT as denoted by arrow ${\rm \uppercase\expandafter{\romannumeral1}}$ by linearly varying the hopping amplitude $w_2$ as  
\be\label{eq:lqp}
w_2(t)=w_2^i+sgn(w_2^f-w_2^i)Rt 
\ee
from time $t_i=0$ to final time $t_f=|w_2^f-w_2^i|/R$.  The system undergoes a  phase transition from  the $N_{xy}=1$ phase to the $N_{xy}=3$ phase.  In the following we will study the scaling behaviors of  number of excitations  with respective to the quenching rate $R$ for  different boundary conditions.   %And the other parameters are fixed as $v=0.4,$  $w_D=0.5,$ $w_1=1,$ and $w_{m>2}=0$, respectively. 
As shown in Fig.~\ref{fig:Bo}, one can see that for different boundary conditions, the number of excitations $N_{exc}$ exhibit a power-law scaling behavior with respect to quenching rate $R$: $N_{exc} \sim R^{\alpha} $
but with distinctive exponents $\alpha$. Specifically, $\alpha=2$ for PBC, while $\alpha=1/2$ for HBC and OBC.

We first discuss the case of HBC with  periodic boundary conditions along $x$ and open boundary conditions along $y$ direction. To understand the physics of exponent $1/2$, one notices that there appears a pair of two-fold degenerate edge-bands in the energy gap, and the wave functions of these edge  states are localized at the boundaries. As one increases $w_2$, the gap of edge-bands  closes at the critical point and then opens again, as shown in Fig.~\ref{fig:Bo}(a-1).  From finite size scaling, one can determine the exponents  $z=1$ and $\nu=1$ \cite{sm}. According to  the KZ mechanism, this transition should lead to an exponent $\alpha= d \nu/(1+z\nu) =1$ if one adopts  the real dimension $d=2$ \cite{Polkovnikov2005,Polkovnikov2008, Sengupta2008}, which is in contrary to our observation of $\alpha=1/2$. This discrepancy can be understood by noting that the effective dimension should be $d^{\rm eff}=1$ rather than $2$, since the phase transition is dominated by  the edge-bands gap closing while the edge states are localized on the $y$-edge and thus are effectively one-dimension with $d^{\rm eff}=1$. To further validate this picture, the post-quench excitations are found to  populate mainly on the edge-bands, as shown in Figs.~\ref{fig:Bo}(a-3) and \ref{fig:Bo}(a-4), in which the pre-quench population and post-quench repopulation of the energy levels are illustrated.  

 % Beyond adiabatic limit with larger quench rate, the excitations of bulk-bands become important, thus disrupting the $1/2$ power-law scaling.  
To analytically understand the exponent $1/2$, we  develop an effective Landau-Zener model that takes into account only the contribution from edge-bands. For simplicity, we consider a system of $C_{2v}$-symmetry with $v_x=v_y=v$, $w_{m>1}=0$, $w_D=0$, and $w_{1,y}>w_{1,x}$. This model  hosts a boundary-obstructed phase transition at $v_c=w_{1,x}$ as the parameter $v$ is varied. When one takes the hybrid boundary conditions with periodic and open along $x$ and $y$ directions, respectively, a pair of two-fold degenerate bands appear in the bulk-energy gap. %To obtain the effective Hamiltonian, we expand the model in the $y$-direction. Thus the model Hamiltonian now is defined on the half-space given by $y>0$ and there exist two edge states whose wave functions localized at the boundary.   
After analytically writing down the wave functions of the two edge states, one can obtain  the effective Hamiltonian $\mathcal{H}_b$ in the Hilbert space spanned by the two edge states.  In SM \cite{sm}, we show that $\mathcal{H}_b$ reproduces  the eigenenergies of the original model Hamiltonian. 
	Next we adopt a quench protocol such that  parameter $v$ is varied with time as $v(t)=g/t$ from time $t_i=0^+$ to infinity.  Here, $g$ characterizes the quench rate. The nonadiabatic  dynamics is dictated by the time-dependent Schr$\mathsf{\ddot{o}}$dinger equation
	$ i\partial_t\vert\psi(t)\rangle=\mathcal{H}_b(t)\vert \psi(t)\rangle$
	with the time-dependent Hamiltonian
	\be\label{eq:bou22}
	\mathcal{H}_b(t)=-(g/t+\cos k_x)\sigma_x-\sin k_x\sigma_y.
	\ee
 The Landau-Zener transition probability $p_{k_x}$ is \cite{Ye2020,Fang2022,Sun2022}: 
	\be\label{eq:pkx12}
	p_{k_x}=\frac{e^{-2\pi g \cos k_x}-e^{-2\pi g}}{e^{2\pi g}-e^{-2\pi g}},
	\ee
from which, one can see that the excitations are mainly produced near $k_x=\pi$. In the adiabatic limit $g \rar \infty$, the expectation value of the density of {edge-bands} excitations becomes
	\be\label{eq:nk22}
	n^e_{exc}=\frac{2}{2\pi}\int_{0}^{2\pi}dk_xp_{k_x} \sim \frac{1}{\pi}\sqrt{\frac{1}{g}},
	\ee
	Here the factor $2$ takes account of the two edges. This effective Landau-Zener model  predicts the same exponent $\alpha=1/2$ with the numerical results in Fig.~\ref{fig:Bo}(a-1), and thus further confirms that the edge-band excitations dominate the KZ behavior of BoOPT.   %A detailed discussion on the comparisons between the total excitations and edge excitations can be found in SM \cite{sm}, which can further confirm that the edge-bands excitations dominate the Kibble-Zurek behavior of BoOPT. 

For OBC, the scaling exponent of excitations is the same as that of HBC, as expected from KZ mechanism. To confirm that the effective dimension is $d^{\rm eff}=1$, we plot the post-quench distribution of excitations in real-space (the inset of Fig.~\ref{fig:Bo}(b-2)). It is seen that the excitations are entirely localized on the edges of the system.
%It is stressed that in calculating the number of excitations $N_{exc}$, we do not consider the contribution of zero-energy models.

However, things become totally different when one considers the PBC, in which case an exponent of $2$ is found, see Fig.~\ref{fig:Bo}(c-2). The exponent $2$ is consistent with the results reported in previous studies for the adiabatic regime \cite{Rams2019, Schmitt2022}, which  means that the excitaions are strongly suppressed in the adiabatic limit. This phenomena can be understood from the energy band which is always gapped even at the critical point of $w_2$. Therefore, the KZ mechanism doesn't apply here, since there is no frozen region during the dynamical process even for sufficiently slow quenching rate. The scaling behavior in this case is sensitive to the quenching protocol.   An exact calculation of excitation based on a four-level Landau-Zener model shows that the excitation is exponentially suppressed at large quenching rate, see SM \cite{sm}. However, in practical numerical calculation,  the quench protocol (\ref{eq:lqp}) has a nonzero initial-time derivative at $t_i=0$, and inevitably produces additional excitation that displays a scaling behavior with exponent $\alpha=2$.  In SM \cite{sm}, we show that an exponent $\alpha=4$ would be obtained if we adopt an alternative smooth protocols with vanishing first-time derivative at initial time.
 
\begin{figure}[t]
	\centering
	\epsfig{file=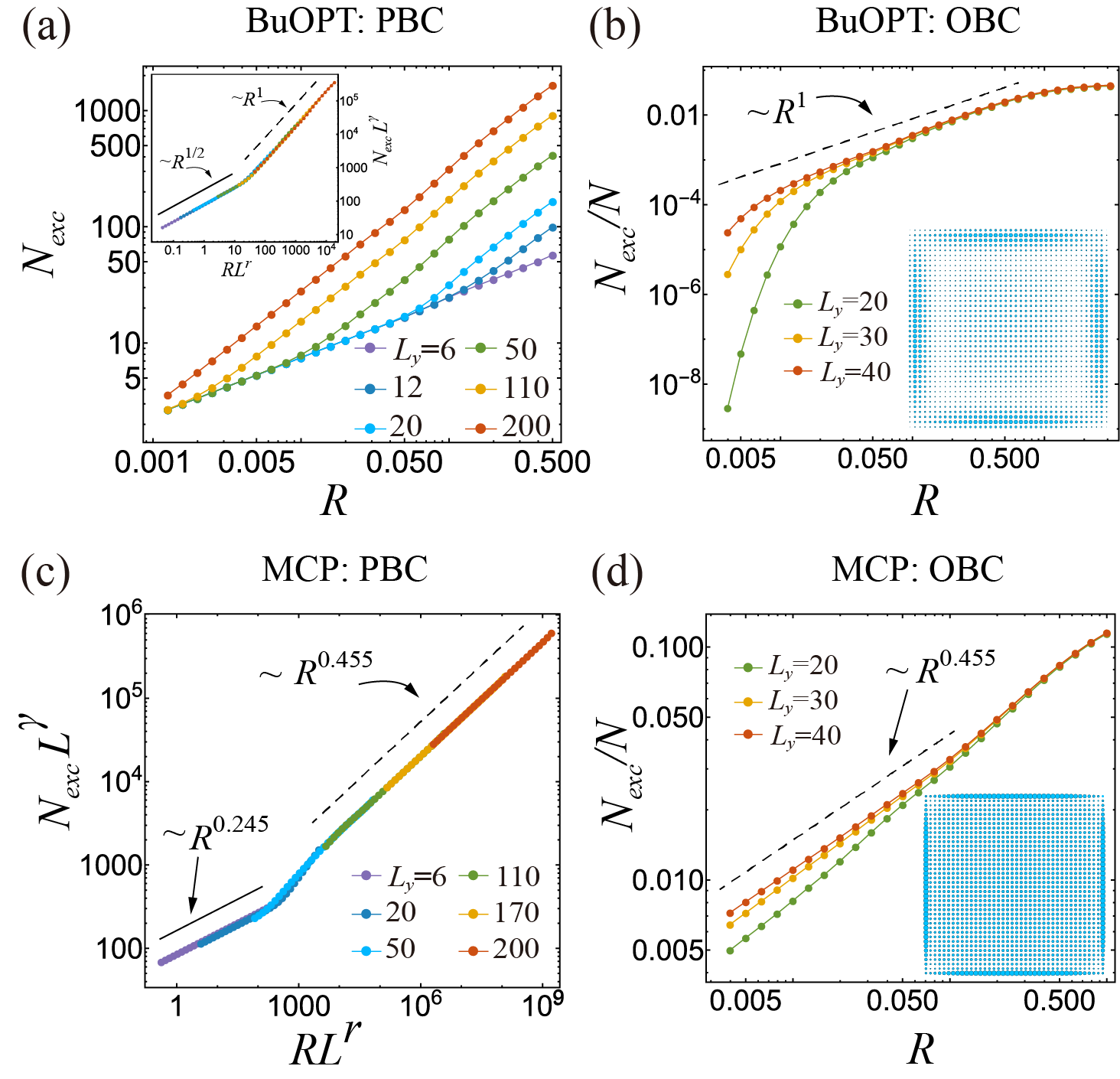, width=3.5in}
	\caption{The number of excitations $N_{exc}$ produced during linear quench across BuOPT with  (a) PBC and (b) OBC for different system size. %Different symbols correspond to different system widths $L_y$. $N_{exc}$ fulfills the KZ scaling $N_{exc}/N\propto R$ as the thermodynamic limit is approached, which is independent of the boundary conditions. 
	Inset of (a) shows that the plots for different $L_y$ collapse to a common scaling according to Eq.~(\ref{eq:rs}). %The black solid and dashed lines represent the $R^{1/2}$ and $R^1$ scaling, respectively. 
	Inset of (b) displays the real-spatial distribution of post-quench excitations. (c)-(d) Same as (a)-(b) but for the quench through the MCP. %The black solid and dashed lines represent the $R^{0.250}$ and $R^{0.455}$ scaling, respectively. In (d) , the black dashed line is a fitted $R^{0.455}$ scaling.
	}
	\label{fig:Bu}
\end{figure}

{\it Quench across BuOPT.}-- For comparison, we proceed to consider the quench across the bulk-obstructed phase transitions with bulk-band gap closing (path ${\rm \uppercase\expandafter{\romannumeral2}}$) . We vary the parameter $w_2$ under the same quench protocol (\ref{eq:lqp}) with $w_2^i=1.3$  ($N_{xy}=3$ phase) to $w_2^f=2.3$ ($N_{xy}=4$ phase). 

As shown in Fig.~\ref{fig:Bu}(a-b), the exponents in the adiabatic limit for different boundary conditions are the same $\alpha=1$, which is in contrast to the case of BoOPT. 
In the inset of Fig.~\ref{fig:Bu}(b), we also plot the post-quench distribution of excitations in real space for the system with OBC. It is seen that the excitations are nearly uniformly distributed in real-space, indicating that the bulk-bands dominate in the phase transition. 

To understand the scaling exponent, we  perform the finite-size scaling according to  KZ scaling hypothesis: 
\be\label{eq:rs}
N_{exc}(L_y,R)=L_y^\gamma F_1(RL_y^r),
\ee
and find that the plots for different size $L_y$ collapse to a single scaling function, as shown in the inset of Fig.~\ref{fig:Bu}(a). Here, $F_1$ is a nonuniversal scaling function and $\gamma$, $r$ are the scaling parameters. Especially $r$ is given by the dynamical exponent $z$ and the exponent of correlation length $\nu$: $r=z+1/\nu$.  From finite-size scaling, we obtain $\gamma = 1$ and $r= 2$, which are consistent with the results  $z=1$ and $\nu=1$ obtained from finite-size scaling on the energy gap (see SM \cite{sm}). 

Different from the result of BoOPT, the exponent $\alpha=1$ in the BuOPT case conforms to the prediction of KZ theory with dimension $d=2$. 
In SM \cite{sm}, we also develop a four-level Landau-Zener problem and present an analytical analysis of this power-law scaling. Note that,  one also observes a regime with $1/2$ exponent in the inset of Fig.~\ref{fig:Bu}(a). This exponent can be attributed to the finite-size effect, since in our calculation the size in $x$ direction $L_x$ is fixed and is set to be large. Therefore, for small size of $L_y$, the system is effectively of dimension $d=1$ \cite{Liou2018}.

{\it Quenches across  the MCP.}--For quench path ${\rm \uppercase\expandafter{\romannumeral3}}$, we vary the  hopping amplitude $v$ : 
$v(t)=v^i+sgn(v^f-v^i)Rt$ from $t_i=0$ to $t_f=|v^f-v^i|/R$ with  $v^i=1.5$ ($N_{xy}=0$ phase) and $v^f=0.5$ ($N_{xy}=3$ phase). Figure~\ref{fig:Bu}(c) shows that the number of excitations scales with quenching rate $R$ with the same exponent $\alpha=0.455$ for different boundary conditions. 

We also perform the finite-size scaling for the PBC case (c) according to Eq.~(\ref{eq:rs}), and the fitting scaling parameters are $\gamma=1.000$ and $r=4.396$, which agrees with the exponents $z=2.420$ and $\nu=0.506$ obtained from finite-size scaling of energy gap, as detailed in SM \cite{sm}. 
These values of exponents predict an scaling exponent $\alpha=0.455$, in consistent with the KZ prediction with $d=2$. Again, due to the finite-size effect, the slope becomes $0.245$ in the slow quench regime for small size system.

 {\it Conclusion.}--We have studied the Kibble-Zurek behavior after a linear quench through the BoOPTs, BuOPTs, and MCP in a chiral and $C_{4v}$-symmetric, long-range QTI model. In the case of the quench through the BoOPTs, one has to take into account the effective dimension of edge state to correctly explain the scaling exponent, and the exponents are totally different for different boundary conditions. This unique Kibble-Zurek behavior across BoOPT is in sharp contrast with the cases of BuOPT and MCP, in which the scaling exponents  are not affected by boundary conditions and can be explained by the real dimension of system. It should be   emphasized  that the underling physics of  effective dimension here is different from the case of a quench through a $(d-m)$-dimensional critical surface \cite{Sengupta2008}, while in our case, only one single critical point is traversed.    Our work reflects the intriguing physics induced by the interplay between topology and nonadiabatic dynamics, and constitutes a significant extension of our current understanding of defect production due to a quench.
 
 {\it Acknowledgements.}--
We thank Adolfo del Campo for helpful discussions. This paper was supported by the National Key Research
and Development Program of the Ministry of Science and Technology (Grant No. 2021YFA1200700), the National Natural Science Foundation of China (Grants No. 12275075), and the Fundamental Research Funds for the Central Universities of China.

%---------------------------------------------------------

\begin{appendix}
	\begin{widetext}
	In this Supplemental Material, we provide more details of numerical and analytical calculations of density of excitations in boundary-obstructed phase transition and bulk-obstructed phase transition with different boundary conditions. We also provide the detailed results of the finite-size scaling.
	\section{Model and Hamiltonian}
Here we provide  a detailed description of the model used in the main text. The more general form of the Hamiltonian is given by: 
	\begin{equation}
	\begin{aligned}
	\hat{\mathcal{H}} =\sum_{i}&\sum_{j}\{v_x(-\hat{c}^\dagger_{i,j}\hat{a}_{i,j}+\hat{d}^\dagger_{i,j}\hat{b}_{i,j})+
	v_y(\hat{b}^\dagger_{i,j}\hat{c}_{i,j}+\hat{a}^\dagger_{i,j}\hat{d}_{i,j})\\
	&+\sum _m^M[w_{mx}(-\hat{c}^\dagger_{i,j}\hat{a}_{i,j+m}+\hat{b}^\dagger_{i,j}\hat{d}_{i,j+m})
	+w_{my}(\hat{d}^\dagger_{i,j}\hat{a}_{i+m,j}+\hat{b}^\dagger_{i,j}\hat{c}_{i+m,j})]\\
	&+w_D(-\hat{a}^\dagger_{i+1,j+1}\hat{d}_{i,j}+\hat{a}^\dagger_{i+1,j+1}\hat{c}_{i,j}
	-\hat{d}^\dagger_{i-1,j+1}\hat{a}_{i,j}-\hat{c}^\dagger_{i+1,j+1}\hat{b}_{i,j}\\
	&+\hat{a}^\dagger_{i-1,j+1}\hat{c}_{i,j}-\hat{c}^\dagger_{i+1,j-1}\hat{b}_{i,j}
	-\hat{d}^\dagger_{i+1,j+1}\hat{b}_{i,j}-\hat{d}^\dagger_{i-1,j+1}\hat{b}_{i,j})+H.c.\}.	
	\end{aligned}
	\label{eq:hr}
	\end{equation}
Here, $\hat{a}^\dagger_{i,j} (\hat{b}^\dagger_{i,j},\hat{c}^\dagger_{i,j},\hat{d}^\dagger_{i,j})$ is the creation operator for degree of freedom $1(2,3,4)$ in unit cell $(i,j)$. $v_x$ and $v_y$ represent amplitudes of hopping within a unit cell. $M$ determines the maximum long-range hopping and $w_{m,x/y}$ represent the amplitudes of hopping to the $m$th nearest-neighbor unit cells in the horizontal and vertical directions, respectively. $w_D$ is the  hopping among nearest-neighbor unit cells along the diagonal directions. The negative signs in Eq.~(\ref{eq:hr}) , represented by the red dashed and solid lines in Fig.~$1$(a) of the main text, come from the gauge choice for a $\pi$ flux which threads through each plaquette of the system. The corresponding Bloch Hamiltonian  reads 
\be\label{eq:hk1}
\mathcal{H}(\mathbf{k})=\mathbf{h}(\mathbf{k})\cdot\mathbf{\Gamma}=\sum_{i=1}^{4}h_i(\mathbf{k})\Gamma_i,
\ee
where $\Gamma_0=\tau_3\otimes\sigma_0,\ \Gamma_{k}=-\tau_1\otimes\sigma_{k},\ \Gamma_4=-\tau_2\otimes\sigma_0$ for $k=1,2,$ and $3$; $\tau, \sigma$ are Pauli matrices, and $\sigma_0$ is the $2 \times 2$ identity matrix. The $\mathbf{h}(\mathbf{k})$ is given by

	\begin{equation}\label{eq:hk2}
	\begin{split}
	&h_1(\mathbf{k})=-v_y-\sum_m^Mw_{m,y}\cos(mk_y)+2w_D \cos k_x\cos k_y;\\
	&h_2(\mathbf{k})=\sum_m^Mw_{m,y}\sin(mk_y)-2w_D \cos k_x\sin k_y;\\
	&h_3(\mathbf{k})=v_x+\sum_m^Mw_{m,x}\cos(mk_x)-2w_D \cos k_y\cos k_x;\\
	&h_4(\mathbf{k})=\sum_m^Mw_{m,x} \sin(mk_x)-2w_D \cos k_y\sin k_x,
	\end{split}
	\end{equation}

The matrix form of the Bloch Hamiltonian is

	\be\label{eq:h2}
	\mathcal{H}(\mathbf{k}) = \left( \begin{array}{cccc}
		0 & 0 & -h_3+ih_4 &  i h_2-h_1  \\
		0 & 0 & -h_1 -ih_2 & h_3+i h_4  \\
		-h_3-ih_4 &  i h_2-h_1 &0 & 0  \\
		-h_1 -ih_2  & h_3-i h_4 & 0 & 0
	\end{array}
	\right).
	\ee

\section{Analytical calculation of the number of excitations under PBC }	
	Here we provide an analytical calculation of the number of excitations under PBC. It should be emphasized that this formalism enables us to consider the excitations both in BoOPT and BuOPT under PBC. Without loss of generality, we set  $v_x=v_y=v(t)$ and vary $v(t)$ to drive the system slowly through quantum phase transitions. To obtain an analytical result, the first step is to put all the time-dependent terms on the diagonal elements. For this purpose, we first use the properties of Dirac matrices to introduce a unitary transformation 
	$\mathbf{T}_1$=$\mathbf{e}^{(\Gamma_1\cdot\Gamma_4-\Gamma_4\cdot\Gamma_1)\pi/8}$ so that the Hamiltonian becomes $\mathcal{H}_1 = \mathbf{T}_1 \mathcal{H} \mathbf{T}_1^{-1}$:

		\be\label{eq:h3}
		\mathcal{H}_1(\mathbf{k})= \left( \begin{array}{cccc}
			0 & 0 & -i h_1-h_3 & i h_2- h_4  \\
			0 & 0 & -i h_2-h_4 & h_3-i h_1  \\
			i h_1-h_3 & i h_2-h_4 &0 & 0  \\
			-i h_2-h_4  & i h_1+h_3 & 0 & 0
		\end{array}
		\right).
		\ee

	Then, a  second unitary transformation $\mathbf{T}_2$ is introduced to make the Hamiltonian look more simpler
	\be\label{eq:t2}
	\mathbf{T}_2 = \left( \begin{array}{cccc}
		0 & \frac{1}{2}-\frac{i}{2} & 0 & \frac{1}{\sqrt{2}}  \\
		-\frac{1}{2}-\frac{i}{2} & 0 & \frac{1}{\sqrt{2}} &0  \\
		0 &-\frac{1}{2}+\frac{i}{2} &0 & \frac{1}{\sqrt{2}}  \\
		\frac{1}{2}+\frac{i}{2}  & 0 & \frac{1}{\sqrt{2}} & 0
	\end{array}
	\right).
	\ee
Making the unitary transformation $\mathcal{H}_2(\mathbf{k})=\mathbf{T}_2 \mathcal{H}_1(\mathbf{k})\mathbf{T}_2^{\dagger}$ and setting the $f_0=\frac{h_3-h_1}{\sqrt{2}}$, $f_1=\frac{h_2-h_4}{\sqrt{2}}$, $f_2=\frac{h_2+h_4}{\sqrt{2}}$, $f_3=\frac{h_3+h_1}{\sqrt{2}}$,
	the Hamiltonian can be written as

		\be\label{h4}
		\mathcal{H}_2(\mathbf{k})= \left( \begin{array}{cccc}
			f_0 & 0 & -i f_3 & -i f_1-f_2  \\
			0 & f_0 & -i f_1+f_2 & i f_3  \\
			i f_3 & i f_1+f_2 &-f_0 & 0  \\
			i f_1-f_2  & -i f_3 & 0 & -f_0
		\end{array}
		\right).
		\ee
	Note that all the time-dependent terms are now on the diagonal elements.

	%Next, we consider a quench in the $C_{4v}$ symmetric system, i.e. $w_x=w_y=w$ and $v_x=v_y=v$. Without loss of generality, we vary the parameter $v$ with time as $v=g/t$. The parameter $g$ determines the quenching rate. It varies from $0$ to $\infty$, corresponding to a continuous crossover from the sudden limit ($g=0$) to an adiabatic limit ($g\rightarrow\infty$).  The form of $g/t$ enables us to quench the system from the initially topological invariant $N_{xy}=0$ regime at $t=0^+$, through the BuOTP at $g/t=w$, to a final $N_{xy}=1$ regime at $t=\infty$ by keeping the other parameters unchanged. 
	In what follows, one can set the $w_{m>1}=0$, $w_D=0$ and $w_{1,y}=nw_{1,x}$. We consider a such quench protocol that $v$ is varied with time as $v(t)=g/t$. The parameter $g$ determines the quenching rate. It varies from $0$ to $\infty$, corresponding to a continuous crossover from the sudden limit ($g=0$) to an adiabatic limit ($g\rightarrow\infty$). The time-dependent Schr$\mathsf{\ddot{o}}$dinger equation can be explicitly written as
	\be\label{eq:s1}
	i\frac{d}{dt}\left( \begin{array}{c}
		c_1\\ c_2\\ c_3\\ c_4   
	\end{array}  
	\right)=\left( \begin{array}{cc}
		\frac{\sqrt{2}g}{t}+f_0' & -i \mathbf{f}\cdot \mathbf{\sigma}   \\
		i \mathbf{f}\cdot \mathbf{ \sigma} & -\frac{\sqrt{2}g}{t}-f_0'
	\end{array}
	\right)\left( \begin{array}{c}
		c_1\\ c_2\\ c_3\\ c_4   
	\end{array}  \right),
	\ee
	where $f_0'= (w_x\cos k_x+w_y\cos k_y)/\sqrt{2}$ and $\mathbf{f}=(f_1,f_2,f_3)$. By noting the identity $(i \mathbf{f}\cdot \mathbf{ \sigma})(-i \mathbf{f}\cdot \mathbf{ \sigma})=f^2$ with $f^2=\sum_{j=1}^{3}f_j^2$, one can decouple the above four equations into two separated ones. First, one can define a new vector $(c_1',c_2')^T=\hat{u}(c_1,c_2)^T$ with $\hat{u}=i\mathbf{f}\cdot\mathbf{\sigma}/f$. Then one can obtain two coupled time-dependent problems in the basis of $(c_1',c_2')^T$ and $(c_3,c_4)^T$:
	
	\be\label{eq:s2}
	i\frac{d}{dt}\left( \begin{array}{c}
		c_1'\\ c_2'   
	\end{array}  
	\right)=\left( \begin{array}{c}
		\frac{\sqrt{2}g}{t}+f_0'
	\end{array}
	\right)\left( \begin{array}{c}
		c_1'\\ c_2'   
	\end{array}  \right)+f\left( \begin{array}{c}
		c_3\\ c_4   
	\end{array}  \right),
	\ee
	
	and
	
	\be\label{eq:s3}
	i\frac{d}{dt}\left( \begin{array}{c}
		c_3\\ c_4   
	\end{array}  
	\right)=-\left( \begin{array}{c}
		\frac{\sqrt{2}g}{t}+f_0'
	\end{array}
	\right)\left( \begin{array}{c}
		c_3\\ c_4   
	\end{array}  \right)+f\left( \begin{array}{c}
		c_1'\\ c_2'   
	\end{array}  \right).
	\ee
	One can see that, now $c_1'$ and $c_3$ are coupled together, but they are decoupled from $c_2'$ and $c_4$. Thus we obtain the two decoupled two-state Landau-Zener (LZ) problem:
	\be\label{eq:s4}
	i\frac{d}{dt}\left( \begin{array}{c}
		a\\ b   
	\end{array}  
	\right)=\left( \begin{array}{cc}
		\frac{\sqrt{2}g}{t}+f_0' & f\\
		f &-\frac{\sqrt{2}g}{t}-f_0'
	\end{array}
	\right)\left( \begin{array}{c}
		a\\ b  
	\end{array}  \right),
	\ee
	whose solution has been obtained in Ref.\cite{Ye2020,Fang2022,Sun2022}. Explicitly, starting from the initial ground state $\left(0,1\right)^T$, the final state vector is written as 
	\be\label{phi1}
	\vert \psi_\mathbf{k}(t)\rangle=\sqrt{p_\mathbf{k}}e^{-i\varepsilon t}\vert+\rangle+\sqrt{1-p_\mathbf{k}} e^{i\varepsilon t +\phi_0}\vert-\rangle,
	\ee
	where $\phi_0$ is a relative phase factor that is independent of time. Here, $\vert\pm\rangle$ are the two instantaneous eigenvectors of the two-level Hamiltonian $H=f_0'\sigma_z+f\sigma_x$, and $p_{\mathbf{k}}$ is the transition probability corresponding to the $\mathbf{k}$ mode:
	\be\label{p1}
	p_{\mathbf{k}}=\frac{e^{-2\sqrt{2}\pi g\mathsf{cos}\theta}-e^{-2\sqrt{2}\pi g}}{e^{2\sqrt{2}\pi g}-e^{-2\sqrt{2}\pi g}}
	\ee
	with $\cos \theta=f_0'/\sqrt{f_0'^2+f^2}=\frac{\left(\cos k_x+n\cos k_y\right)}{\sqrt{2}\sqrt{1+n^2}}$.
By the LZ formula (\ref{p1}) we can calculate the number of excitations as:
	\be\label{n1}
	N_{exc}=2\sum_{\mathbf{k}}p_{\mathbf{k}},
	\ee
	where factor 2 comes from that Eq.~(\ref{eq:s1}) is composed of two decoupled two-state LZ models. 
	
	When considering the system with $C_{4v}$-symmetry, i.e. $n=1$, the  $v(t)= g/t$ quench protocol enables us to quench the system from the initially topological invariant $N_{xy}=0$ regime at $t_i=0^+$, through the BuOPT at $t_c=g/w_{1,x}$, to a final $N_{xy}=1$ regime at $t_f=\infty$. For slow quench with $g\gg 1$, the $e^{-2\sqrt{2}\pi g}$ terms in Eq.~(\ref{p1}) can be ignored. In addition, only modes near $\mathbf{k}=\left(\pi, \pi\right)$ can get excited. For these modes we can use the LZ formula [\ref{p1}] for excitation probability:
	\be\label{p2}
	p_{\mathbf{k}}\simeq e^{-\frac{\sqrt{2}}{2}\pi g\left[ (k_x-\pi)^2+(k_y-\pi)^2\right] }.
	\ee
	In the thermodynamic limit, the sum in Eq.~(\ref{n1}) can be replaced by an integral. The expectation value of the density of excitations becomes
	\be\label{n2}
	n_{exc}=\lim_{N\to\infty}\frac{N_{exc}}{N}=\frac{1}{2\pi^2}\int_{0}^{2\pi}d\mathbf{k}p_{\mathbf{k}}=\frac{\sqrt{2}}{2\pi^2g},
	\ee
	where $N$ is the number of unit cells of the system. The density scales with $g^{-1}$ in agreement with KZM. As depicted in Fig.~\ref{fig:anpbc}(a), the agreement of this expression (\ref{n2}) with results of numerical is remarkable.

	When setting the $n\neq1$, the system is $C_{2v}$-symmetric and BoOPT appear at $v=w_{1,x}$ $(v=w_{1,y})$ for $\vert n\vert>1$  $(\vert n\vert<1)$. Likewise, let $v$ varying with time as $g/t$ and the transition probability $p_{\mathbf{k}}$ for $\mathbf{k}$ mode can be calculated by Eq.(\ref{p1}). We approximate (\ref{p1}) up to the second order in $k_x$ and  $k_y$. Thus the transition probability $p_{\mathbf{k}}$ can be written as
	\be\label{p}
	p_{\mathbf{k}}\simeq e^{2\pi g(\frac{1+n}{\sqrt{1+n^2}}-\sqrt{2})} e^{-\frac{\pi g}{\sqrt{1+n^2}}\left[ (k_x-\pi)^2+n(k_y-\pi)^2\right] }.
	\ee
	In the thermodynamic limit, the sum in Eq.~(\ref{n1}) can be replaced by an integral. The expectation value of the density of excitations becomes
	\be\label{n}
	n_{exc}=\lim_{N\to\infty}\frac{N_{exc}}{N}= \frac{1}{2\pi^2}\int_{0}^{2\pi}d\mathbf{k}p_{\mathbf{k}}=e^{2\pi g(\frac{1+n}{\sqrt{1+n^2}}-\sqrt{2})}\frac{1}{2\pi^2g}\sqrt{\frac{1+n^2}{n}}.
	\ee
	The density of  excitations exhibit an exponential decay shown in Fig.~\ref{fig:anpbc}(b) (black solid line). The exponential decay means the excitations are strongly suppressed in the adiabatic regime, due to the absence of gap closing at critical point. When setting $n=1$, Eq.~(\ref{n}) returns to (\ref{n2}).

	Interestingly, the excitations obtained by numerical simulations are much larger than the theoretical values in the large $g$ region and decay with $g$ like $g^{-2}$ rather than exponentially, see Fig.~\ref{fig:anpbc}(b) (red six-pointed star). This discrepancy stems from the fact that the excitations is mainly generated by the discontinuity of quench velocity at the staring point in the case. Besides, a detailed study on quench dynamics in $C_{2v}$-symmetric system can be found in section G.

\begin{figure}[htbp]
	\centering
	\epsfig{file=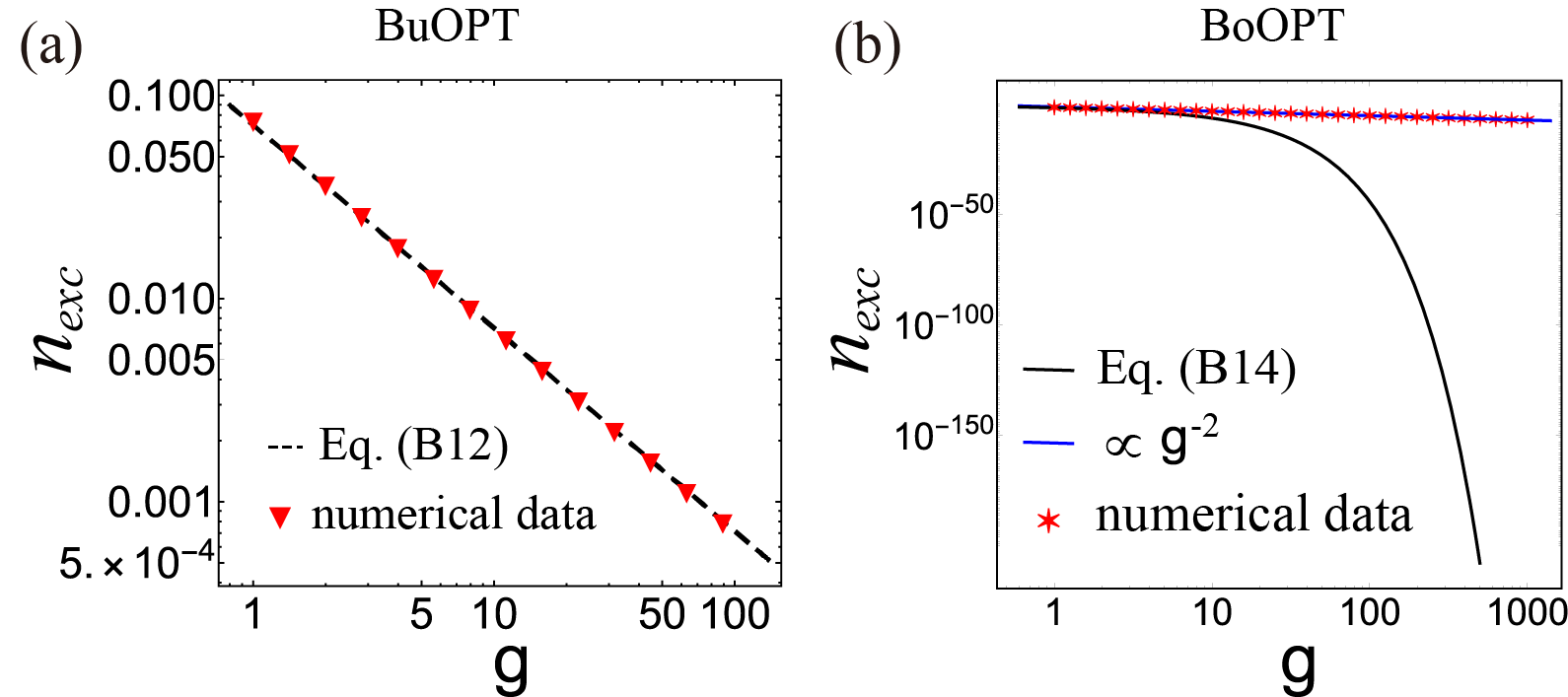, width=3.5in}
	\caption{Comparison of numerical data with analytical results for PBC. (a) $n=1$, the quench process passes through BuOPT. The density of excitations in the function of quench time $g$. Here, the numerical data (symbols) agree with the analytical result of Eq.~(\ref{n2}) (dashed). (b) $n=3$, the quench process passes through BoOPT. The resulting density of excitations is shown with red symbol and much larger than the theoretical prediction (\ref{n}) (black solid line). The fit exhibits power-law scaling with exponent $-2$ (blue solid line).}
	\label{fig:anpbc}
\end{figure}

	\section{Analytical calculation of the number of edge-band excitations under HBC }
	
In this section, we derive the number of edge-band excitations generated during the quench across the BoOPTs and show that the edge-band excitations dominate the power-law scaling. Similarly, we consider a $C_{2v}$-symmetric system with $w_{1,y}=nw_{1,x}$. Without loss of generality, we set the $\vert n\vert>1$ and a pair of two-fold degenerate bands appear in the bulk-energy gap for the system with periodic and open boundary conditions along $x$ and $y$ directions, respectively.  Due to the periodic boundary conditions along $x$ direction, it is convenient to use the basis $\vert k_x\rangle \otimes \vert y \rangle$, where $k_x\in [0,2\pi)$ is Bloch wave vector and $y\in \{1,...,L_y\}$ is a lattice site in $y$ direction. In this basis, the Hamiltonian is block diagonal, $\mathcal{H}=\sum_{k_x}\vert k_x\rangle\langle k_x\vert \otimes \mathcal{H}(k_x)$, with 
	
		\begin{equation}\label{eq:hkx}
		\begin{split}
	\mathcal{H}(k_x)&=\sum_{y=1}^{L_y-1}\vert y+1\rangle\langle y\vert \otimes \frac{-w_y}{2}(\Gamma_1-i\Gamma_2) +H.c.\\
		&+\sum_{y=1}^{L_y}\vert y\rangle\langle y\vert\otimes\{-v\Gamma_1+(v+w_x\cos k_x)\Gamma_3 \\&+w_x\sin k_x\Gamma_4\}.
		\end{split}
		\end{equation}
	
	An electron occupying the $m$th subband with eigenenergy $\varepsilon_m(k_x)$ is described by the wave function $\vert\Psi_m(k_x)\rangle=\vert k_x\rangle\otimes\vert u_m(k_y)\rangle$, where $\vert u_m(k_x)\rangle$ is an eigenstate of $\mathcal{H}(k_x)$. The energy levels as a function of the amplitude of hopping $v$ are shown in Fig.~\ref{fig:anhbc}(a). One can see that the edge-band gap closing occurs at critical value $v_c$ .
	
	For computational convenience, we expand the model in the $y$-direction, so the model Hamiltonian is defined on the half-space given by $y>0$. Then the lattice model in the tight binding approximation can be mapped into a continuous model. One can replace :
	
	\begin{equation}\label{eq:nsrp1}
	\begin{aligned}
	&\sin k_y \rightarrow k_y,\\
	&\cos k_y\rightarrow 1-\frac{1}{2}k_y^2.
	\end{aligned}
	\end{equation}
	And the Bloch Hamiltonian Eq.~(\ref{eq:hk1}) can be divide into two parts
	\begin{align}
	&\mathcal{H}(\mathbf{k})=\mathcal{H}_x+\mathcal{H}_y,\label{eq:div1}\\
	&\mathcal{H}_x=-v \Gamma_1+(v+w_x\cos k_x)\Gamma_3+w_x\sin k_x \Gamma_4,\label{eq:div2}\\
	&\mathcal{H}_y=w_y(\frac{1}{2}k_y^2-1)\Gamma_1+w_y k_y \Gamma_2.\label{eq:div3}
	\end{align}
	All $k_y$ dependent terms are included in $\mathcal{H}_y$. We replace $k_y$ by $-i\partial_y$ and obtain the eigenvalue equation
	\be\label{eq:enq1}
	\mathcal{H}_y(k_y\rightarrow-i\partial_y)\Psi(y)=E\Psi(y).
	\ee
	Since $\Gamma_1=-\tau_1\otimes\sigma_1$ and $\Gamma_2=-\tau_1\otimes\sigma_2$ are both block diagonal, the Hamiltonian $\mathcal{H}_y$ is also block and the eigenstates have the form
	
	\textbf{\begin{equation}\label{eq:psi1}
		\begin{split}
		\vert\Psi_1\rangle=\left( \begin{array}{c}
		\psi_0\\ \mathbf{0}   
		\end{array}  
		\right),\vert\Psi_2\rangle=\left( \begin{array}{c}
		\mathbf{0}\\\psi_0
		\end{array}
		\right),
		\end{split}
		\end{equation}}
	where $\mathbf{0}$ is a two-component zero vector. To obtain the edge states, the wave function $\psi_0(y)$ should be localized at the edge and satisfies the eigenequation
	\be \label{eq:enq2}
	[w_y(\frac{1}{2}\partial^2_{y}+1)\sigma_1+iw_y\partial_y\sigma_2]\psi_0(y)=E\psi_0(y).
	\ee
	The eigenequation (\ref{eq:enq2}) exhibits the chiral symmetry; therefore we expect that special bound states with $E=0$ can exist. With the wave function ansatz $\psi_0=\phi e^{\lambda y}$, the above equation can be simplified as 
	\be\label{eq:enq3}
	(\frac{1}{2}\lambda^2+1)\sigma_3\phi=-\lambda \phi.
	\ee
	The two-component wave function $\phi$ should be the eigenstate of the Pauli matrix $\sigma_3$. Let us define $\sigma_3\phi_\eta=\eta\phi_{\eta} (\eta=\pm1)$, one can obtained the equation
	\be\label{eq:enq4}
	\frac{1}{2}\lambda^2+\eta \lambda+1=0.
	\ee
	The two roots satisfy the relation $\lambda_1+\lambda_2=-2\eta$ and $\lambda_1\lambda_2=2$. We require that the wave function vanishes at $y=0$ and $y=+\infty$:
	\be\label{eq:bcs}
	\psi_0(y=0)=\psi_0(y=+\infty)=0.
	\ee
	So the two roots should be negative, and only $\phi_{+1}$ satisfies the boundary condition for a bound state. Thus $\lambda_{1,2}$ satisfy:
	\be\label{eq:enq5}
	\lambda_{1,2}=-1\pm i.
	\ee
	Thus the wave function for the bound states is given by

		\be\label{eq:psi2}
		\psi_0(y)=C\left( \begin{array}{c}
			1\\ 0   
		\end{array}\right) (e^{\lambda_1y}-e^{\lambda_2y})=2i Ce^{-y}\sin y \left( \begin{array}{c}
			1\\ 0   
		\end{array}\right),
		\ee
	where $C$ is the normalization constant. The $x$-dependent part $\mathcal{H}_x$ (\ref{eq:div2}) is regarded as the perturbation to the one-dimensional Hamiltonian. In this way, we have a one-dimensional effective model in the subspace $\{\vert\Psi_1\rangle, \vert\Psi_2\rangle\}$ for the edge states:

		\begin{equation}\label{eq:bou}
		\begin{split}
	\mathcal{H}_b(k_x)=&\left( \langle\Psi_1\vert, \langle \Psi_2\vert\right) \mathcal{H}_x\left( \begin{array}{c}
		\vert\Psi_1\rangle\\ \vert\Psi_2\rangle   
		\end{array}\right).\\
		=&-\left( \begin{array}{cc}
		0 & v+w_xe^{-ik_x}\\
		v+w_xe^{ik_x} &0.
		\end{array}
		\right)
		\end{split}
		\end{equation}
The effective Hamiltonian $\mathcal{H}_b$ corresponds to a 1D Su-Schrieffer-Heeger (SSH) model. The energy levels of $\mathcal{H}_b$ as a function of $v$ are shown with a red dashed line in Fig.~\ref{fig:anhbc}(a). One can see that $\mathcal{H}_b$ recreates the exact eigenenergies well. 
	
	Next, we calculate the number of excitations of these edge-bands in the slow-quench regime by the two-bands Hamiltonian $\mathcal{H}_b$. Just as the quench protocol used in the above, we vary parameter $v$ with time as $v(t)=g/t$. The dynamics is dictated by the time-dependent Schr$\mathsf{\ddot{o}}$dinger equation,
	\be\label{eq:seq}
	i\partial_t\vert\psi(t)\rangle=\mathcal{H}_b(t)\vert \psi(t)\rangle,
	\ee
	with the time-dependent Hamiltonian
	\be\label{eq:bou2}
	\mathcal{H}_b(t)=-(\frac{g}{t}+\cos k_x)\sigma_x-\sin k_x\sigma_y,
	\ee
	 The Landau-Zener transition probability $p_{k_x}$ is
	\be\label{eq:pkx1}
	p_{k_x}=\frac{e^{-2\pi g \cos k_x}-e^{-2\pi g}}{e^{2\pi g}-e^{-2\pi g}}.
	\ee
	As the excitations are for slow quenches $(g\gg 0)$ generated only at momenta $k_x$ close to $\pi$, we approximate the Eq.~(\ref{eq:pkx1}) up to the second order in $k_x$. Thus the 
	transition probability $p_{k_x}$ can be written as 
	\be\label{eq:pkx2}
	p_{k_x}=e^{-\pi g (k_x-\pi)^2}.
	\ee
	This analytical result is consistent with the numerical results, as shown in Fig.~\ref{fig:anhbc}(b). With the LZ formula (\ref{eq:pkx2}) we can calculate the number of edge-bands excitations as :
	\be\label{eq:nk1}
	N_{exc}^e=2\sum_{k_x}p_{k_x},
	\ee
	where factor 2 follows from the fact that our real system has two boundaries. In the thermodynamic limit, the sum in Eq.~(\ref{eq:nk1}) can be replaced by an integral. The expectation value of the density of excitations becomes
	\be\label{eq:nk2}
	n^e_{exc}=\lim_{L_x\to\infty}\frac{N^g_{exc}}{L_x}=\frac{1}{\pi}\int_{0}^{2\pi}dk_xp_{k_x}=\frac{1}{\pi}\sqrt{\frac{1}{g}},
	\ee
	where $L_x$ is the number of unit cells in the $x$ direction. The density scales with $g^{-1/2}$ in agreement with KZM with effective dimension $d^{\rm eff}=1$. The agreement of the expression (\ref{eq:nk2}) with numerical results is shown in Fig.~\ref{fig:anhbc}(c). It is worth noting that the edge-band excitations $N^e_{exc}$ and total excitations $N_{exc}$ are essentially equal, which suggests that the former dominates the power-law scaling of excitations in the case of BoOPTs.
	
%\begin{figure}[htbp]
	%\centering
	%\epsfig{file=s2.png, width=1.75in}
	%\caption{Passing through the BoOPT with a smooth quench protocol in the system with PBC. The number of excitations in function of quench rate $R$ for different system widths $L_y$. For smooth protocol (\ref{eq:sqp}) with no discontinuity of velocity at the beginning of the quench, the number of excitations increases as $R^4$. Dashed lines provide the guidance for an eye for the expected scaling (slop). }
	%\label{fig:lsqp}
%\end{figure}
	
\begin{figure*}[htbp]
	\centering
	\epsfig{file=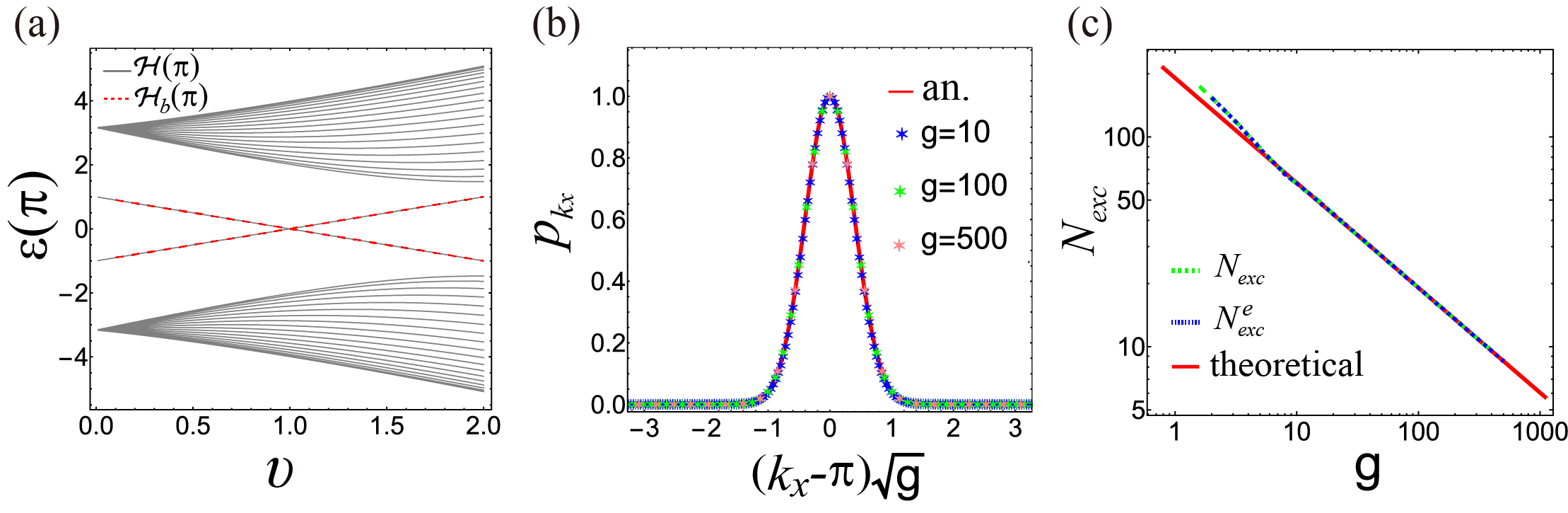, width=5.25in}
	\caption{ (a) Energy levels of the whole Hamiltonian $\mathcal{H}(k_x)$ (gray) and of the two-level Hamiltonian $\mathcal{H}_b(k_x)$ (red dashed) at $k_x=\pi$ in function of hopping amplitude $v$. (b) Momentum distribution of edge-bands excitations in a strip with $L_y=20$ after quenches with $g=10$ (blue), $g=100$ (green), $g=500$ (pink). The rescaled analytical result of Eq.~(\ref{eq:pkx2}) is shown by the red line. (c) The numerical results of number of excitations in function of $g$. The total excitations $N_{exc}$ (green dashed) and edge-band excitations $N^e_{exc}$ (blue dashed) have the same value and agree with the theoretical prediction (\ref{eq:nk2}) multiplied by $L_x$ (red solid line), which indicates that the edge-band excitations dominate the power-law scaling in the case of BoOPT. Here $L_x$ is the $x$-direction size taken in our numerical calculations.}
	\label{fig:anhbc}
\end{figure*}

\section{ Passing BoOPTs with smooth quench protocols under PBC}
	In this section, we investigate the effect of quench protocols as the system is driven through BoOPTs under PBC. In the main text, we considered a linear quench protocol and showed that the number of excitations scales with the quench rate $R$ as $R^2$ (see Fig.~2(c-2) of main text). For comparison, we also discuss a common smooth quench protocol. In the protocol, we vary the amplitude of hoping $w_2$ as
	\be\label{eq:sqp}
	w_2(t)=w_{2c}+Rt-\frac{16}{27}(Rt)^3,
	\ee
	between initial $t_i=-\frac{3}{4R}$ and final $t_f=\frac{3}{4R}$, where $w_{2c}$ is the critical value. The protocol has a vanishing first-time derivative at $t_i$ ($t_f$) due to the cubic term, which limits the generation of additional excitations at the start of the quench. In the vicinity of the critical point around $t\approx 0$, the protocol exhibits the same slope as Eq.~(5) of the main text. The resulting number of excitations is plotted in Fig.~\ref{fig:hbc}(a), which shows that excitations increase with $R$ like $R^4$ rather than $R^2$.

	%From Eq.~(3) of the main text, it is seen that the first time derivative of the protocol vanishes at $t_i$, which limits the generation of additional excitations at the start of the quench. To distinguish the above result from the excitations generated by a nonanalytic temporal behavior at the starting $t_i$, we consider a linear quench 
	%\be\label{eq:q1}
	%w_2(t)=w_2^i+(w_2^f-w_2^i)\sin^2(Rt/2),
	%\ee
	%starting at $t_i=0$, crossing the BoOPTs, and ending at $t_f=\vert w_2^f-w_2^i\vert/R$. The resulting number of excitations scales as $R^2$, shown in Fig.~\ref{fig:lsqp}(a), which is consistent with the results reported in previous articles for the adiabatic regime. 

\section{Numerical results for HBC and the summary of the power-law scaling }
In the case of quench across BuOPT or MCP, we only presented the results of PBC and OBC in Fig.~3 of the main text. For completeness, we show the results of HBC in this section. In Fig.~\ref{fig:hbc}(b), we present the number of excitations produced during the quench across BuOPT as the thermodynamic limit is approached. It readily follows that the number of excitations is accurately described by the scaling $N_{exc}/N\propto R$, which is in complete agreement with results of FBC and OBC. The resulting excitations during quench across the MCP is plot in Fig.~\ref{fig:hbc}(c). One can see that the number of excitations scales as $N_{exc}/N\propto R^{0.455}$, which is independent of boundary conditions.   Besides, power-law scalings of the number of excitaions for different types of phase transition under different boundary conditions are summarized in Table~\ref{table:1}.

\begin{figure}[htbp]
	\centering
	\epsfig{file=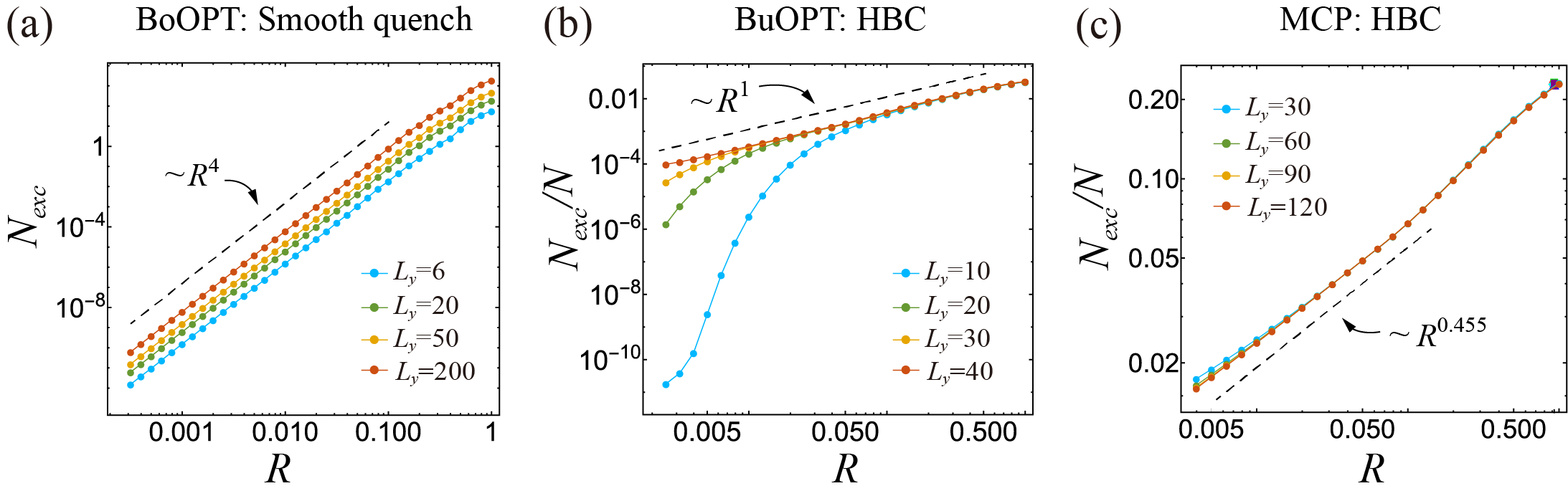, width=5.25in}
	\caption{(a) Passing through the BoOPT with a smooth quench protocol in the system with PBC. The number of excitations in function of quench rate $R$ for different system widths $L_y$. For smooth protocol (\ref{eq:sqp}) with no discontinuity of velocity at the beginning of the quench, the number of excitations increases as $R^4$. The results of HBC for a linear quench across (b) the BuoPT, and (c) MCP. $N_{exc}$ fulfills the KZ scaling $N_{exc}/N\propto R$ and $R^{0.455}$, respectively, as the thermodynamic limit is approached. $N$ is the number of occupied states before quench. }
	\label{fig:hbc}
\end{figure}	

\begin{table}[htpb]
	\caption{The number of excitations exhibits a power-law scaling with quench rate $R$ as $N_{exc}\sim R^{\alpha}$. The $\alpha$ obtained in the different cases is summarized as follows. The results in the first line correspond to a quench across BoOPT and different boundary conditions have been considered. The other two lines contain results corresponding to quenches across BuOPT and MCP, respectively.}
	\centering
	\begin{tabular}{llll}
		\toprule[0.8pt]
		\midrule[0.8pt]
		Types of phase&  \multirow{2}{*}{\quad\quad\quad PBC}& \multirow{2}{*}{\quad\quad\quad HBC}&  \multirow{2}{*}{\quad\quad\quad OBC}\\
		transition & & &\\
		\midrule[0.8pt]
		BoOPTs &\quad\quad\quad $4$ &\quad\quad\quad $0.5$ &\quad\quad\quad $0.5$\\

		BuOPTs &\quad\quad\quad $1$ &\quad\quad\quad $1$ &\quad\quad\quad $1$\\
		
		MCP &\quad\quad\quad $0.455$ &\quad\quad\quad $0.455$ &\quad\quad\quad $0.455$\\
		\midrule[0.8pt]
		\bottomrule[0.8pt]	
		\label{table:1}
	\end{tabular}	
\end{table}

	\section{ Comparison of total excitations with edge excitations}
	
Utilizing the real-space distribution of excitation $n_{exc}(\br)=\sum_{c, v}|\la \br|\Psi_c \ra|^2 \vert\langle\Psi_c\vert\Phi_\nu\rangle\vert^2$, one can define the number of edge excitations in the real-space as
\be\label{eq:nr}
N_{exc}^{e}=\sum_{\mathbf{r}\in S'}n_{exc}(\mathbf{r}) ,
\ee
where $S'$ denotes the edge region of the system (shaded area in Fig.~1(a) of the main text). Here we compare the number of total excitations $N_{exc}$ with edge excitations $N^e_{exc}$ and show that the edge excitations determine the scaling of excitations in the case of BoOPT. Figure~\ref{fig:cp}(a) illustrates the numerical results of excitations after a linear quench across BoOPT for OBC and one can see that the total excitations and edge excitations are almost identical in the (relatively) slow-quench regime. Thus, for the BoOPT, the excitations are dominated by the contributions from the edges.  This results emphasize that for BoOPT, the effective dimension should be taken as $d^{\rm eff}=1$ instead of the actual dimension of the system $d=2$. In the fast-quench regime, the larger quench rate induces additional bulk excitations, giving rise to deviations from the edge excitations.  For the cases of both BuOPT and MCP, however, the bulk-excitations cannot be ignored and the scaling of excitations is no longer determined by edge excitations, as shown in Figs.~\ref{fig:cp}(b) and \ref{fig:cp}(c).
	\begin{figure*}[htbp]
	\centering
	\epsfig{file=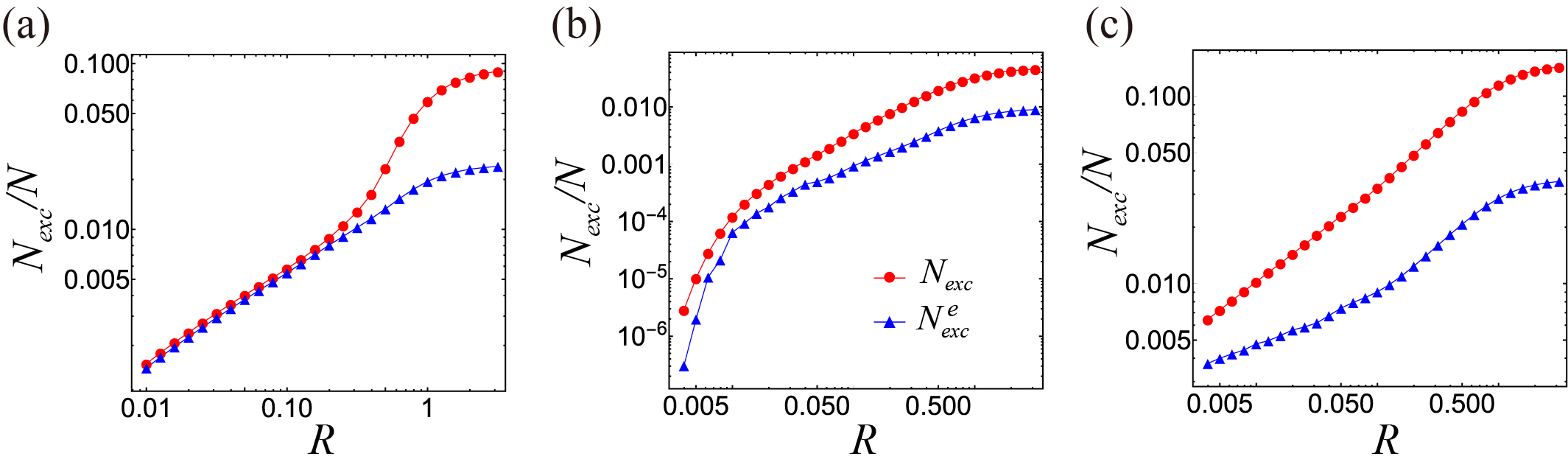, width=5.25in}
	\caption{Comparison of total excitations $N_{exc}$ with edge excitations $N^e_{exc}$ defined in Eq.~(\ref{eq:nr}) for quench through (a) BoOPT, (b) BuOPT, and (c) MCP in the system with OBC. For BoOPT (a), the total excitations and edge excitations are almost identical in the slow-quench regime, which indicates that the power-law scaling is determined by the excitations from edge. In (b) and (c), the scaling of excitations is no longer determined by edge excitations.}
	\label{fig:cp}
	\end{figure*}

	\section{ Obtaining the critical exponents by using the finite-size scaling on the energy gap } 
The energy gap goes as $\Delta E \propto \vert \epsilon\vert^{z\nu}$, close to the critical point, where $z$ and $\nu$ are the universal dynamical and correlation length critical exponents, respectively. For finite-size systems, the finite-size effects appear in the vicinity of the critical point. And near criticality, the dependence of the energy gap $\Delta E$ with the finite linear system sizes $L$ is given by
	\be\label{eq:srs2}
	\Delta E(L,u)=L^{-z}f_E(|u-u_c|L^{1/\nu}),
	\ee
	 here $u$ can be any control parameter of the system and $u_c$ is the corresponding critical value. And $f_E$ is a nonuniversal scaling function. To obtain critical exponents, we change the coefficients $\mu_{1,2}$ until we observe the collapse of $\Delta EL^{\mu_1}$ as a function of $|u-u_c|L^{\mu_2}$ for all used linear system sizes $L$. In this way, we extract $\mu_1=z$ and $\nu=1/\mu_2$. In Fig.~\ref{fig:ce}, we show results for the energy gap between the ground state and the first excited state in the vicinity of phase transitions as a function of the amplitude of hopping for different types of phase transitions. It is worth emphasizing that in calculating the energy gap, we do not consider the occupation of zero-energy models. The result of the finite-size scaling (FSS) of energy gap $\Delta E$ for BoOPT is depicted in the inset of Fig.~\ref{fig:ce}(a). From the scaling, we obtain the critical exponents $z=0.985$ and $\nu=0.980$. In insets of Figs.~\ref{fig:ce}(b) and \ref{fig:ce}(c), we show the rescaling results of energy gap for BuOPT and MCP and obtain the corresponding critical exponents $z=0.995$, $\nu=0.971$ and $z=2.420$, $\nu=0.506$, respectively. Our estimates for critical exponents $z$ and $\nu$ obtained using FSS on the energy gap for different types of phase transitions are summarized in Table~\ref{table:2}.
\begin{figure*}[htbp]
	\centering
	\epsfig{file=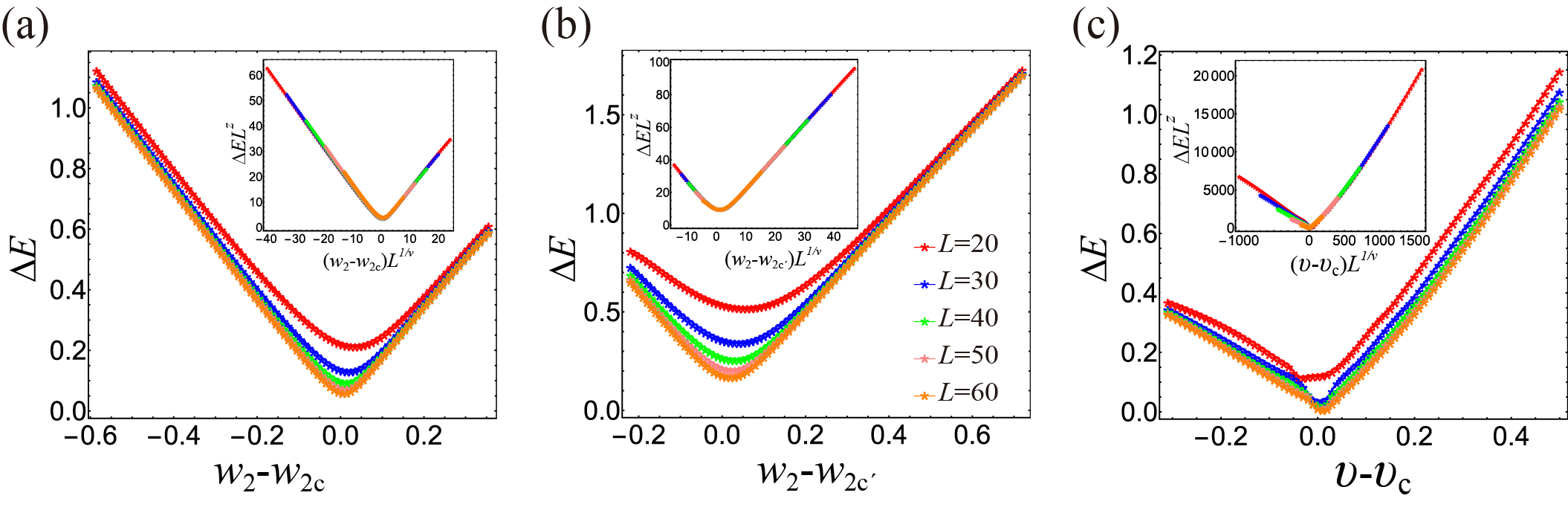, width=5.25in}
	\caption{Energy gap as a function of the amplitudes of hopping in the vicinity of (a) BoOPT, (b) BuOPT, and (c) MCP in a full open $C_{4v}$-symmetric QTI model with $L_x=L_y=L$. Insets show the collapse of the data after finite-size rescaling with the critical exponents (a) $z=0.985$, $\nu=0.980$, (b) $z=0.995$, $\nu=0.971$, and (c) $z=2.420$, $\nu=0.506$, respectively.   }
	\label{fig:ce}
\end{figure*}	

\begin{table}[h]
	\caption{The critical exponents for different types of phase transitions. The critical exponents are extracted from the FSS on the energy gap for the system with OBCs. The system size is $L_x*L_y$ and we take the $L_x=L_y=L\in\{20,30,40,50,60\}$ rungs. }
	\centering
	\begin{tabular}{lll}
		\toprule[0.8pt]
		\midrule[0.8pt]
		Types of phase&  \multirow{2}{*}{\quad\quad\quad\quad\quad\quad\quad z}& \multirow{2}{*}{\quad\quad\quad\quad\quad\quad\quad $\nu$}\\
		transition & &\\
		\midrule[0.8pt]
		BoOPTs &\quad\quad\quad\quad\quad\quad\quad $0.985$ &\quad\quad\quad\quad\quad\quad\quad $0.980$\\
		
		BuOPTs &\quad\quad\quad\quad\quad\quad\quad $0.995$ &\quad\quad\quad\quad\quad\quad\quad $0.971$ \\
		
		MCP &\quad\quad\quad\quad\quad\quad\quad $2.420$ &\quad\quad\quad\quad\quad\quad\quad $0.506$ \\
		\midrule[0.8pt]
		\bottomrule[0.8pt]
		\label{table:2}
	\end{tabular}
	
\end{table}

\begin{figure*}[htbp]
	\centering
	\epsfig{file=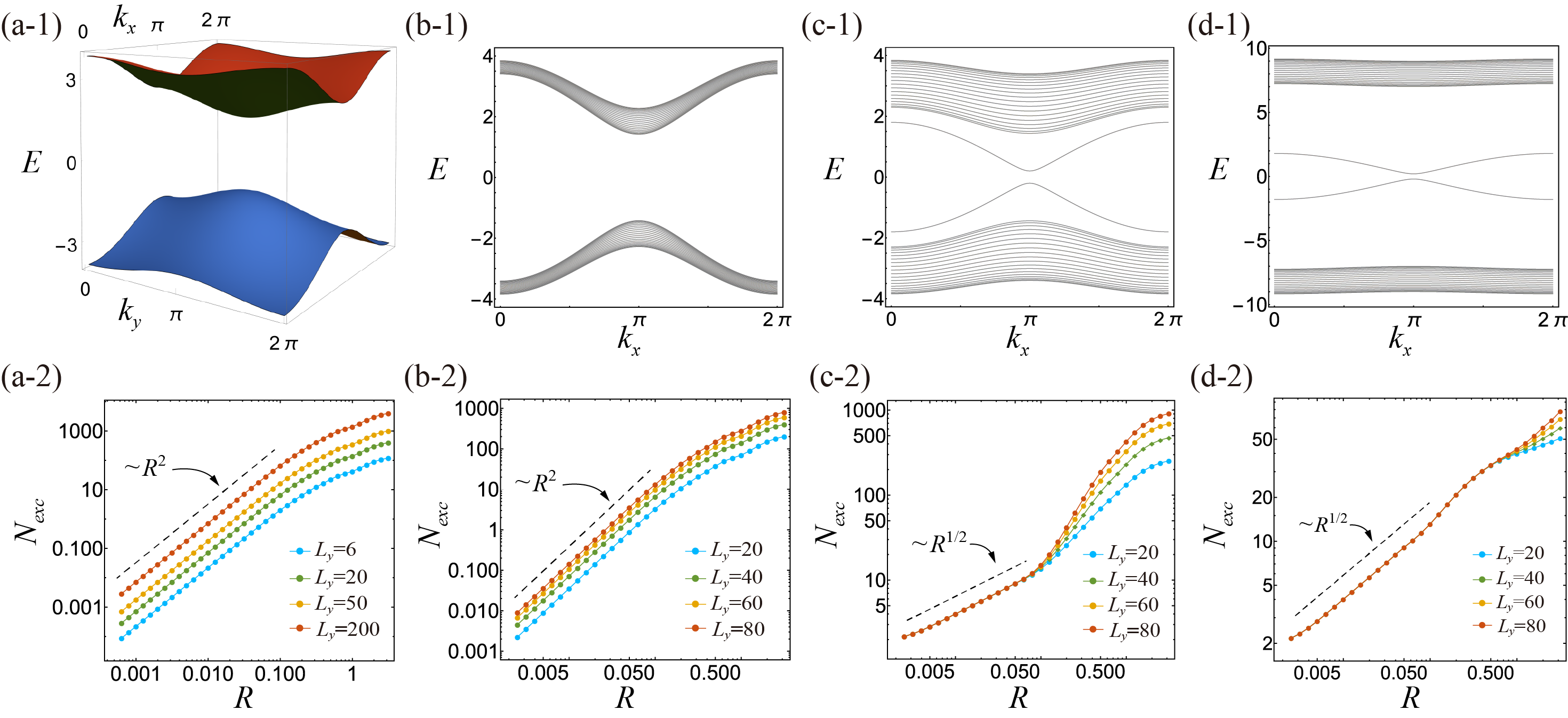, width=7in}
	\caption{Quench results 
		for the $C_{2v}$-symmetric system. PBC: (a-1) Band structure at the critical point. Each energy band is twofold degenerated for a total of four bands. (a-2) Scaling of the number of excitation produced during a linear quench across the critical point. The black dashed line represent $R^2$ scaling. HBC: (1) Dispersions for different $n$ for the $C_{2v}$-symmetric QTI model with open (periodic) boundary conditions along the (b) $x$ ($y$), $n=3$, (c) $y$ ($x$), $n=3$, and (d) $y$ ($x$) direction, $n=10$. (2) Number of excitations in functions of $R$ for different system widths $L_y$. For relatively slow $R$, the function is a power-law with an exponent close to (b-2) 2, (c-2) 1/2 and (d-2) 1/2 (dashed line). It should be noted that for a lager $n$, the scaling $N_{exc}\propto R^{1/2}$ holds for a larger range of $R$.}
	\label{fig:nshbc}
\end{figure*}

\section{ Kibble-Zurek behavior of the $C_{2v}$-symmetric system}	
In the interest of completeness, we provide a detailed study of the Kibble-Zurek behavior of a $C_{2v}$-symmetric system with $w_{1,y}=nw_{1,x}$ ($\vert n\vert>1$). Same as in the main text, we linearly vary the $w_{1,x}$ with time as

%\be\label{eq:sq1}
%w_{1,x}(t)=w_{1,x}^i+sgn(w_{1,x}^f-w_{1,x}^i)Rt,
%\ee

	\be\label{eq:q1}
w_{1,x}(t)=w_{1,x}^i+sgn(w_{1,x}^f-w_{1,x}^i)Rt,
\ee 
from the trivial regime at $w_{1,x}^i=0.5$, across BoOPT to the topological regime at $w_{1,x}^f=1.5$. For the $C_{2v}$-symmetric system with PBC, there is no energy gap closing at the critical point $v=w_{1,x}$, as shown in Fig.~\ref{fig:nshbc}(a-1). After quenching across the critical point, the resulting number of excitations exhibits a power-law scaling with quench rate $R$ with scaling exponent $\alpha=2$ for various sizes $L_y$, shown in the Fig.~\ref{fig:nshbc}(a-2), which is consistent with Fig.~2(c-2) of the main text.

 For HBC, the band structure of the $C_{2v}$-symmetric system is related to the direction of the open boundary condition. Only if open boundary condition is imposed along the $y$ direction, a pair of two-fold degenerate bands appear in the bulk-energy gap, see Figs.~\ref{fig:nshbc}(b-1) and \ref{fig:nshbc}(c-1). The numerical results of excitations for different boundary conditions are shown in Figs.~\ref{fig:nshbc}(2). For the system with open boundary condition along $x$ direction, the number of excitations scales as $N_{exc}\propto R^2$, which is consistent with the result of PBC. When open boundary condition is imposed along the $y$ direction, however, the number of excitations scales as $N_{exc}\propto R^{1/2}$ in the slow-quench regime. Although beyond adiabatic limit with larger quench rate, the excitations of bulk-bands become numerous, thus disrupting the power-law scaling. Besides the $n$ also affects the band structure. Fixing other parameters, the dispersions corresponding to $n=3$ and $n=10$ are shown in Figs.~\ref{fig:nshbc}(b-1) and \ref{fig:nshbc}(c-1), respectively. It is seen that the gap between bulk bands increases with $n$. Therefore, one can reduce the effects of bulk excitations by increasing $n$, so that the $N_{exc}\propto R^{1/2}$ holds for a larger range of $R$, see Fig.~\ref{fig:nshbc}(c-2).

 \begin{figure}[htbp]
 	\centering
 	\epsfig{file=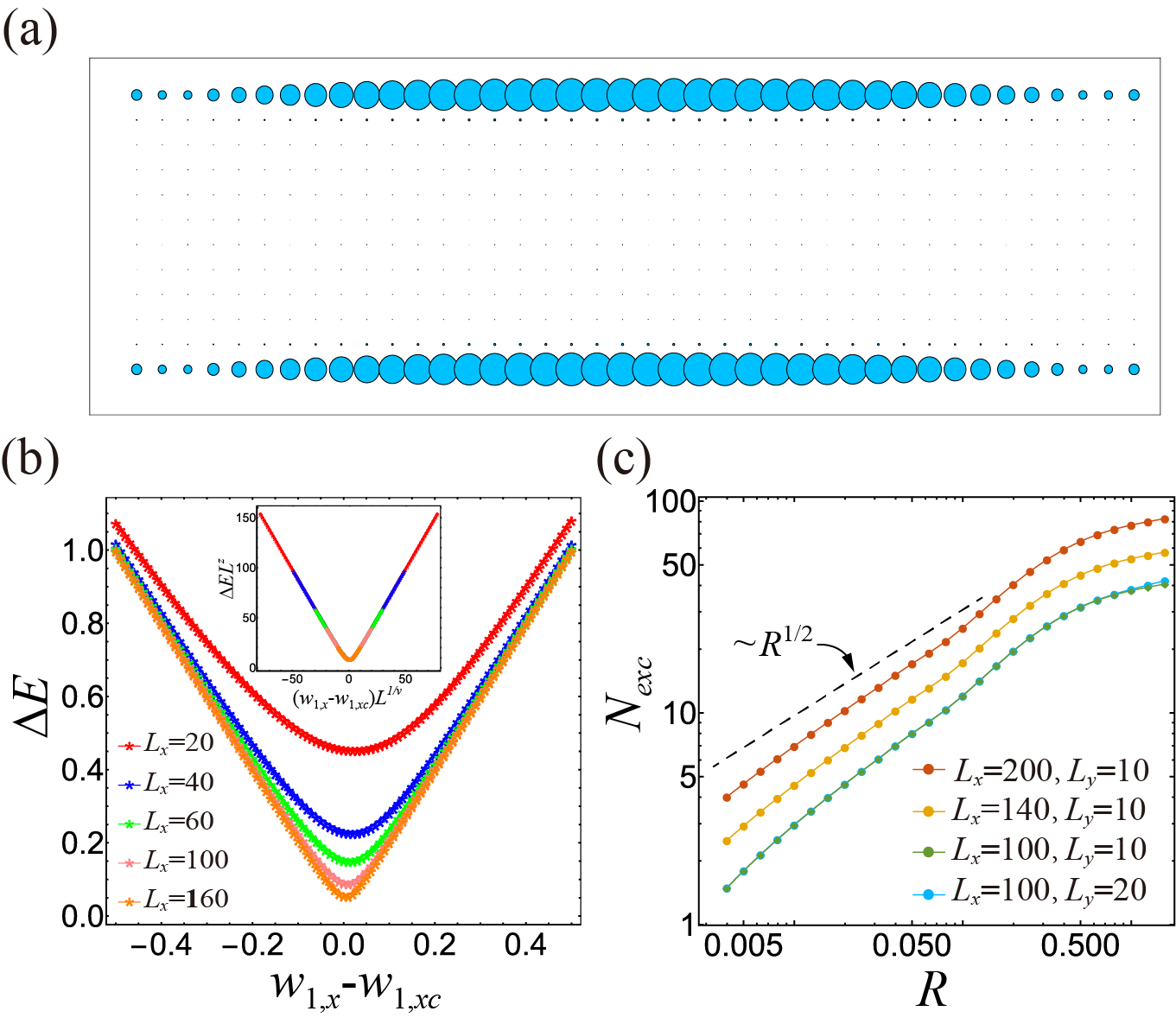, width=3.5in}
 	\caption{The $C_{2v}$-symmetric system with OBC: (a) Spatial distribution of excitations after quench with $R=10^{-1.2}$. (b) Energy gap as function of the amplitude of hopping $w_{1,x}$ in the vicinity of critical point $w_{1,xc}$ for different system lengths $L_x$. The inset shows the collapse of the data after finite-size rescaling with the critical exponents $z=0.995$, $\nu=1.008$. (c) Number of excitations $N_{exc}$ in function of $R$. In the thermodynamic limit, the number of excitations fulfills the scaling $N_{exc}\propto R^{1/2}$. To reduce the effect of finite size, we set the $L_x\gg L_y$.}
 	\label{fig:nsobc}
 \end{figure}
 
Figure~\ref{fig:nsobc}(a) shows that the excitations are entirely localized at the $y$ edges for OBC. To reduce the effect of finite size, we consider strip geometry with $L_x$ sites in the $x$ direction and $L_y$ sites in the $y$ direction and set  $L_x\gg L_y$. The results of excitations for different sizes are shown in Fig.~\ref{fig:nsobc}(c). It readily follows that the number of excitations $N_{exc}$ is described by the scaling $N_{exc}\propto R^{1/2}$ as the thermodynamics limit is approached. Taking into account that $z\nu=1.003$ obtained using FSS on energy gap (see Fig.~\ref{fig:nsobc}(b)), the result conforms to the KZ prediction with effective dimension $d^{\rm eff}=1$.

		\end{widetext}
\end{appendix}
\end{document}